\documentclass{aa}  

\usepackage{graphicx}
\usepackage{txfonts}
\usepackage{float}
\usepackage{caption}
\usepackage{subcaption}
\usepackage{xcolor}

\definecolor{orange}{rgb}{1,0.5,0}

\definecolor{boh}{rgb}{1,0,0}

\begin{document}

   \title{Galaxy populations of ProtoClusters in cosmological hydrodynamical simulations}

   \author{Michela Esposito
          \inst{1,2,3,4}\fnmsep\thanks{michela.esposito@inaf.it}
          ,
          Stefano Borgani\inst{1,2,3,4,5}
          ,
          Veronica Strazzullo\inst{2,3}
          ,
          Maurilio Pannella\inst{2,3}
          ,
          Gian Luigi Granato\inst{2,3,6}
          ,
          Cinthia Ragone-Figueroa\inst{6,2}
          ,
          Alex Saro\inst{1,2,3,4,5}
          ,
          Mario Nonino\inst{2}
          ,
          Milena Valentini\inst{1,2,3,4,5}
          }

   \institute{Department of Physics, University of Trieste,
               via G. Tiepolo 11, I-34131 Trieste, Italy\\
              \email{michela.esposito@phd.units.it}
         \and
             INAF - Observatory of Trieste, via G. Tiepolo 11, I-34143 Trieste, Italy\\
             \email{michela.esposito@inaf.it}
        \and
             IFPU - Institute for Fundamental Physics of the Universe, via Beirut 2, I-34014 Trieste, Italy
        \and
             INFN - Istituto Nazionale di Fisica Nucleare, via Valerio 2, I-34127, Trieste, Italy 
        \and
            ICSC - Italian Research Center on High Performance Computing, Big Data and Quantum Computing, via Magnanelli 2, 40033, Casalecchio di Reno, Italy
        \and
            IATE - Instituto de Astronom\'ia Te\'orica y Experimental, Consejo Nacional de Investigaciones Cient\'ificas y T\'ecnicas de la Rep\'ublica Argentina (CONICET), Universidad Nacional de C\'ordoba, Laprida 854, X5000BGR, C\'ordoba, Argentina\
             }

   \date{Received MMM DDD, YYY; accepted MMM DDD, YYY}

 
  \abstract
   { The study of protoclusters at cosmic noon is essential for understanding the impact on galaxy properties of the environment and of the transformational processes occurring during this epoch.}
   {This work tests the predictions on galaxy evolution of the {\tt DIANOGA} cosmological hydrodynamical simulations of cluster progenitors at $z=2.2$, comparing them with observations, and investigates the environmental effects on galaxy populations by comparing simulated protoclusters with an average volume of the Universe.}
   {We analyze 14 {\tt DIANOGA} protoclusters and a cosmological box of 49  cMpc/h per side, simulated with {\tt OpenGADGET3}. We compare predictions and observations of the galaxy stellar mass function (GSMF), the star-forming Main Sequence (MS), the fraction of star-forming gas, gas depletion times, and the fraction of quenched galaxies. We also compute the rest-frame UV-to-NIR colors of galaxies with the SKIRT-9 radiation transfer code, to analyze UVJ diagrams.}
   {We showed that the {\tt DIANOGA} simulations produce a GSMF in broad agreement with observations. The simulated GSMF shows a higher fraction of high-mass galaxies ($ \rm M_{\ast}>10^{10} \  M_{\odot}$) in massive halos in protoclusters, compared to the cosmological box. The same signal, albeit with lower significance, is also observed in the wide-field protocluster structures, indicating an accelerated evolution of galaxies before their infall into massive halos. Our simulations underestimate star formation rates of galaxies both in protoclusters and in the cosmological box, compared to observed counterparts, due to an underestimation of the star-forming gas reservoirs. We find a weak suppression of SFRs in protocluster galaxies ($\sim0.05$ dex), compared to the cosmological box, increasing up to $\sim0.25$ dex in massive halos, reflecting suppressed cold gas reservoirs. The quenched galaxy fraction varies significantly across different  protocluster halos, consistent with observations at  $z\sim2-2.5$. The simulations show a strong dependence of quenched fractions on host halo mass, as well as an excess of quenched galaxies in the wide-field protocluster region outside of the most massive halos, in comparison with the cosmological box. UVJ diagram analysis shows qualitative agreement with observed color distributions of star-forming and quenched galaxies, except for few massive galaxies in the cores of massive halos where age-dependent extinction results in steeper reddening vectors than typically assumed in observational studies.}
   {}

   \titlerunning{Galaxies in simulated protoclusters}
   \authorrunning{Michela Esposito et al.}
   

   \maketitle
%

\section{Introduction}
Protoclusters (PCs) are regions of the Universe that are expected to collapse into a galaxy cluster by redshift $z=0$ \citep[e.g.,][]{overzier}.  The study of PCs is crucial for understanding the early stages of galaxy cluster formation.
Investigating cluster progenitors at cosmic noon provides an opportunity to examine galaxy populations during an epoch when star formation, Active Galactic Nuclei (AGN) activity, and mergers occur at their highest rates - though with a significant halo-to-halo variation -  at the intersection of gas-rich filaments. These regions thus offer invaluable insights into the interplay of these processes and both the local and global environment in shaping galaxy evolution.
\newline
However, identifying PCs in the high-redshift Universe is challenging. Detecting galaxy overdensities at high redshift does not straightforwardly confirm their future evolution into clusters, as overdensity or halo mass growth predictions can only be made statistically \citep[e.g.,][]{angulo12}. Furthermore, regions that have not yet reached a sufficiently high density may be overlooked, introducing a bias toward denser systems that may not represent the full diversity of protocluster environments.
\newline
Few examples of high-$z$ clusters have been identified through the detection of their diffuse hot intra-cluster gas, either via extended X-ray emission or Sunyaev–Zeldovich effect  \citep[e.g.,][]{santos11,willis20,mantz20}. These cluster selection approaches are (to first order) independent of cluster galaxy population properties.
Another promising and unbiased technique for detecting PCs involves absorption features imprinted by the dense gas of a PC on the spectrum of a UV-bright background source. This method has evolved into a systematic approach to obtain a 3D map of the intergalactic medium permeating the cosmic web, with a resolution of a few megaparsecs, known as Lyman-$\alpha$ forest tomography \citep{lee14,cai16,cai17,zhong21}.
\newline
 On the other hand, the identification of PCs in observational studies has predominantly relied on the detection of galaxy overdensities, with many studies focusing on specific galaxy populations as tracers. The most reliable approach involves spectroscopic surveys, which provide accurate redshifts and robust confirmation of overdensities \citep[e.g.,][]{cucciati18, lemaux18}. However, due to the high cost and time required for extensive spectroscopic observations over large areas, many more PCs (and PC candidates) have been identified using photometric surveys, either by identifying photo-z overdensities \citep[e.g.,][]{eisenhardt08,scoville13,thongkham24} or looking for overdensities of specific galaxy types, such as Lyman Break Galaxies (LBGs), Lyman-$\alpha$ and H$\alpha$ emitters, passive galaxies, distant red galaxies (DRGs), red sequence galaxies, dusty star-forming galaxies (DSFGs) or IRAC color selected sources \citep[e.g.,][]{ steidel00,andreon09,papovich10,spitler12,clements14, strazzullo15,wang16,toshikawa16,daddi17,greenslade18,guaita20,ito23}. The latter methods, while effective, are subject to biases related to the galaxy populations used as tracers, which might affect interpretations and conclusions concerning galaxy evolution in PCs.
Another common technique involves the use of ``biased tracers'', searching for PCs around extreme high-redshift galaxies associated with massive, forming systems. These tracers include high-redshift radio galaxies, quasars, and sub-mm galaxies, which are thought to be preferentially located in protocluster environments \citep[e.g.,][]{pentericci01,hatch11, wylezalek13,wylezalek13b,miller18,koyama21}.
\newline
Over the past three decades since the detection of the first PC, these diverse methods have painted a heterogeneous picture of the early stages of galaxy cluster formation, revealing a vast diversity in cluster progenitor environments at high redshift. These structures allow us to trace the evolution of cluster galaxy populations back to epochs before the collapse of the first massive clusters at $z\lesssim2$, and thus to investigate the nature and impact at early times of different kinds of environmental effects that result in environmental signatures on cluster galaxy populations at later epochs. Observations of cluster galaxies up to $z\sim1$ show that cluster environments are ubiquitously characterized by a population of massive galaxies with a higher fraction of quiescent sources compared to average field regions \citep[e.g.,][]{haines15,vanderburg18,vanderburg20}. This implies that galaxy populations in cluster progenitor environments undergo an accelerated evolution compared to their field counterparts at early times. Consequently, PCs have been sites of enhanced star formation at some point in their evolution, reversing the trend observed at $z\lesssim 1$ \citep[e.g.,][]{oteo18,miller18,smail24}. 
\newline
In this context, several studies have investigated star formation rates (SFRs) in individual PC galaxies compared to field counterparts, looking for environmental signatures on the relation between SFR and stellar mass of star-forming galaxies, known as the ``Main Sequence'' \citep[MS, e.g.,][]{noeske07,elbaz11}. Some of these studies have reported enhanced star formation in galaxies within PCs at cosmic noon \citep[e.g.,][]{wang16,shimakawa18a,monson21,perezmartinez24,staab24}. However, other studies have found that SFRs in PCs are consistent with the field MS at the same redshift  \citep[e.g.,][]{cucciati14,koyama21,polletta21,shi21,perezmartinez23}. 
\newline
Irrespective of an SFR enhancement at the time of observation, PCs may have been sites of more intense star formation at earlier times. The galaxy stellar mass function (GSMF) provides key insight into this, as it encapsulates the integrated star formation history of galaxies over time. Recent studies of some PCs at $z\sim2-3$ have identified an excess of high-mass galaxies in dense environments, compared to the coeval field  \citep{sun24, forrest24}, suggesting that these environments may have undergone an enhanced star formation phase in the past. In contrast, other work has found no significant deviation from the field GSMF in other PCs at $z\sim2-2.5$ \citep[e.g.,][]{edward24}, indicating that their star formation history might not differ greatly from that of the field.
\newline
Another key area of investigation is the fraction of quenched galaxies in PCs, which provides insight into the efficiency of quenching mechanisms in these early stages of cluster assembly. While some studies have found that the fraction of quenched galaxies in high-$z$ PCs is similar to that in the field \citep[e.g.,][]{wang16,forrest24,edward24}, overdensities of passive galaxies have been observed up to $z\sim 4$ \citep[e.g.,][]{zavala19,willis20,mcconachie22,ito23,tanaka24}.
\newline
The limited statistics and observational limitations, combined with differences in the selection of PCs, complicate the understanding of galaxy evolution in cluster progenitors.
In this context, theoretical models in the form of cosmological hydrodynamical simulations can contribute to improving our picture of galaxy evolution, accounting for gas dynamics and astrophysical processes self-consistently in a cosmological context. However, these models need to be validated through a detailed comparison with observations before they can be used as predictive tools. In the very first attempts comparing results from cosmological hydrodynamical simulations to observations \citep[e.g.,][]{saro09}, simulated PCs were predicted to be strongly star-forming.  On the other hand, reproducing observations of galaxy populations in high-$z$ PCs  in detail  has been later found to be challenging for simulations \citep[e.g.,][]{granato15,bassini20,lim21,remus23}. One of the main inconsistencies lies in the underestimation of SFRs in simulated star-forming galaxies at cosmic noon, irrespective of their environment \citep[e.g.,][]{bassini20,akins22,andrews24,ragone24}, which also results in lower integrated SFRs than those observed in PC environments. Recently, \cite{lim24} showed that the FLAMINGO simulations can partially reconcile this discrepancy, with the most massive PC cores in the 1 cGpc and 2.8 cGpc boxes of the FLAMINGO flagship runs matching the observed integrated SFRs at $z\sim2-3$. On the other hand, we reiterate that the underestimation of the star-forming MS in simulations is not limited to the PC environment; it also affects cosmological boxes representing ``average'' regions of the Universe, also referred to as the ``field'' in observational studies. Thus, not only are simulations struggling to match the extreme SFRs reported for a possibly biased population of the most star-forming PCs, but also the field-level star formation at $z\sim2-4$. This indicates a deeper inadequacy of star formation and feedback models in capturing the details of galaxy formation and evolution. Despite these limitations, simulations can still provide valuable insights into the trends of galaxy population properties with their environment, even though the detailed star formation history is not yet well reproduced.
\newline
In this work, we test the predictions of the {\tt DIANOGA} re-simulations of galaxy clusters at $z=2.2$ by comparing simulated galaxy populations with observations at $2\leq z\leq2.5$. The outline of this article is as follows: in Sect. \ref{sec:methods}, we describe the {\tt DIANOGA} set of simulations, the operational definition of PCs adopted in this work, and the radiation transfer simulations performed to model photometric classification of galaxy populations routinely adopted in observational studies. In Sect. \ref{sec:results}, we compare the simulated and observed properties of the galaxy populations in PCs, including the GSMF, the star-forming MS, the fraction of cold star-forming gas, gas depletion times, and the fraction of quenched galaxies. We compare these observables in different environments, to highlight the environmental signatures on the galaxy populations predicted by our simulations. In Sect. \ref{sec:conclusions}, we summarize our main findings. We adopt a \cite{chabrier} initial mass function (IMF), and give magnitudes in the AB system.


\section{Methods}
\label{sec:methods}
\subsection{Simulations}
\label{s:simul}
The analysis presented in this paper is based on a set of 14 simulated galaxy clusters, extracted from the {\tt DIANOGA} set of zoom-in cosmological hydrodynamical simulations \citep[e.g.][]{bonafede11,rasia15}. The adopted cosmology is a flat $ \rm \Lambda CDM$, with $\rm h=0.72$, $\rm \Omega_m=0.24$, $ \rm \Omega_b=0.04$ for the density parameters\footnote{We note that the adopted cosmology differs from the fiducial flat $\Lambda$CDM and the standard Planck 2018 \citep{planck} cosmologies adopted in recent observational studies. However, we verified that the impact of cosmology on halo masses, stellar masses and radii is well within typical statistical and systematic uncertainties in current observations.} associated to total and baryonic matter, $\rm n_s=0.96$ for the primordial spectral index and $ \rm \sigma_8 = 0.8$ for the normalization of the linear power spectrum. The initial conditions have been generated with the zoomed-in initial conditions technique \citep[][]{tormen97}: cluster-size halos of dark matter (DM) particles are identified at $z=0$ in a $1024^3$ DM particles box of 1$\,\rm{h}^{-1}$cGpc per side, then traced back to their positions in the initial conditions at $z\sim180$; the regions defined in this way are re-simulated at higher resolution than the one adopted in the parent simulation, adding gas particles with an initial mass chosen so as to reproduce the cosmic baryon fraction of the assumed cosmology. The re-simulations were performed with a state-of-the-art developer version of the TreePM-Smoothed Particle Hydrodynamics (SPH) code {\tt OpenGADGET3} (\citealt{groth23, damiano24}; evolution of {\tt GADGET-3} and {\tt GADGET-2}, \citealt{springel05b}). Two out of fourteen {\tt DIANOGA} regions
were stopped at $z=1$. The mass resolution for DM particles and the initial mass of gas particles are $\rm m_{DM}= 3.4 \times 10^7 \ \text{h}^{-1} \text{M}_{\odot}$ and  $\rm m_{gas}= 6.2 \times 10^6 \  h^{-1} \text{M}_{\odot}$, respectively. The Plummer-equivalent gravitational softening lengths in the high-resolution regions at $z=0$ are
$\rm \epsilon_{DM}= 1 \ h^{-1} \text{kpc} $, $\rm \epsilon_{gas}= 1 \ h^{-1} \text{kpc}$, and $ \rm \epsilon_{\ast,BH}= 0.25 \ h^{-1} \text{kpc}$, for DM, gas, and star and black hole (BH) particles respectively. All softenings are kept fixed in comoving units, except $\rm \epsilon_{DM}$, held constant in physical units below $z=2$, and fixed in comoving units for $z>2$.
\newline
Our simulations include metal-dependent radiative cooling with a uniform UV background \citep{wiersma09}, star formation out of a multi-phase interstellar medium (ISM), and galactic outflows driven by supernova feedback \citep[ SH03 hereafter]{springel03}. According to the adopted star formation model, gas particles whose density exceeds a threshold density, equivalent to a number density of hydrogen atoms $\rm n_H>0.1$ cm$^{-3}$ are tagged as star-forming multi-phase particles. In such gas particles, a cold and a hot phase coexist in pressure equilibrium, with the cold phase representing the reservoir for star formation. As such, multi-phase particles become eligible to spawn a collisionless star particle stochastically. We assume that each gas particle can generate four generations of star particles so that the typical mass of the latter is $ \rm m_{*}= 1.5 \times 10^6 \ h^{-1} \text{M}_{\odot}$. 
\newline
Stellar evolution and the subsequent chemical enrichment is modeled according to the original implementation presented by \cite{tornatore07}. The model of stellar evolution assumes a \citet{chabrier} IMF; therefore, when comparing with observations in Sect. \ref{sec:results}, observed stellar mass and SFR estimates are converted to this IMF where necessary.
\newline
Our simulations also include the effect of AGN feedback from gas accretion onto supermassive BHs \citep{springel05a}. The gas accretion is described by a Bondi-like, Eddington-limited prescription, also distinguishing between cold and hot gas accretion efficiencies, as described in  \cite{steinborn15}.  The parameters regulating accretion onto BHs and the associated feedback have been tuned to reproduce the observed relation between BH mass and stellar mass of galaxies in the local Universe \citep[][]{magorrian}, using the relation from \cite{mcconell13} and, for Brightest Cluster Galaxies (BCGs), data from \cite{gaspari19}. \citet{bassini20} verified that the $z=0$ galaxy stellar mass function of a lower-resolution (by a factor of 2.5 in mass) version of the { \tt DIANOGA } simulations, including a galaxy formation model quite similar to the one adopted here, is in good agreement with observational results. For the simulations presented here, a forthcoming work (Borgani et al., in prep.) will show a detailed assessment of the agreement between galaxy population properties in this version of the {\tt DIANOGA} simulations and in observations in the nearby Universe.
\newline
In the simulations, galaxies have been identified with the {\tt SubFind} algorithm \citep{springel01,dolag09}.
We use {\tt SubFind} catalogues to define the centers of galaxies and the stars or gas particles bound to them. To compare to galaxy properties as measured from observations, we impose a fixed 2D aperture of 1 arcsec (approximately 8.5 kpc at $z=2.2$)  in radius, along a random line of sight, for consistency with observations typically adopting similar apertures (we compare stellar masses and SFRs computed within these apertures to those directly provided by {\tt SubFind} in Appendix \ref{sec:app1}). To ensure an adequate numerical resolution, we consider galaxies with $ \rm M_{\ast} \geq 10^9 \ M_{\odot}$ in all the analyses presented in Sect. \ref{sec:results}. 
\newline
After identifying the star particles that belong to each galaxy, we compute the corresponding star formation rate (SFR) as follows. Each star particle is characterized by an age (time elapsed since it has been generated). We then compute SFRs averaged over a typical time scale (that is chosen following the observational data to compare with, see Sect. \ref{sec:ms}), by considering the total initial mass of star particles younger than the given time scale and dividing by it, within the adopted aperture. This allows for a more proper comparison with observations, rather than relying on more noisy instantaneous SFRs that could be calculated by summing over the star formation rates of all the star-forming gas particles that {\tt SubFind} associates to the galaxy. We show the comparison between the instantaneous and averaged SFRs in Appendix \ref{sec:app2}.

\subsection{Identification of protoclusters}
\label{sec:pcid}
\begin{figure*}
 \centering
 \includegraphics[width=0.8\textwidth]{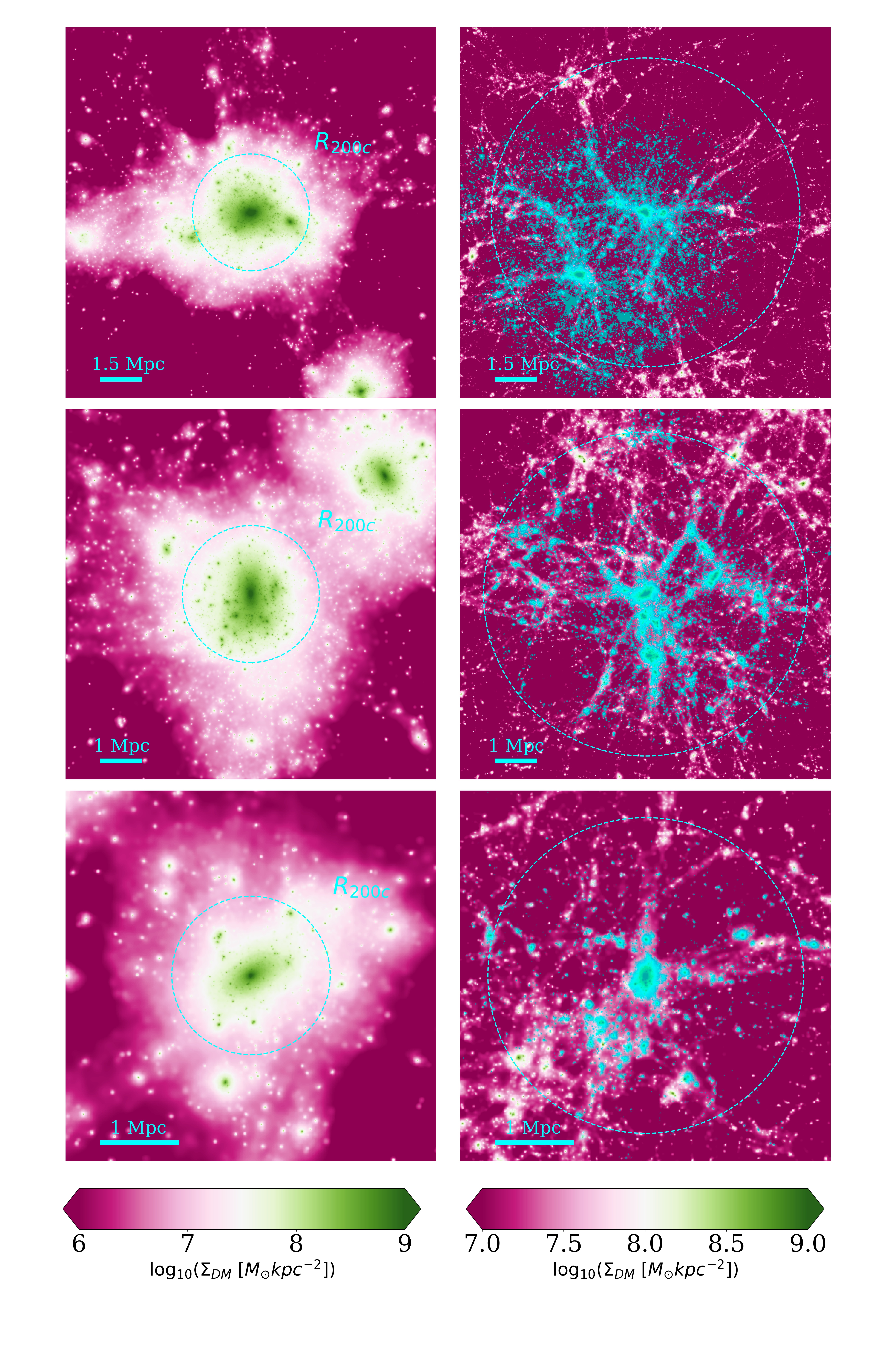}
 \caption{Projected Dark Matter density maps of three simulated clusters at $z=0$ (left column), with total masses $\rm M_{200c}=2 \times 10^{15} \ M_{\odot}$ (top), $ \rm 9\times10^{14} \ M_{\odot}$ (middle) and $ \rm 2\times10^{14} \ M_{\odot}$ (bottom), and their $z=2.2$ progenitors (right column). The cyan circle at $z=0$ traces the $ \rm R_{200c}$ radius of each cluster, the one at $z=2.2$ defines the PC region as defined in the text. DM particles at $z=2.2$ that collapse within the $ \rm R_{200c}$ radius by $z=0$ are plotted in cyan. Densities are computed in pixels of size 15 kpc in the top panel and 10 kpc in the middle and bottom panels.}
 \label{fig:dmmap}
\end{figure*}

\begin{figure}
    \centering
    \includegraphics[width=\linewidth]{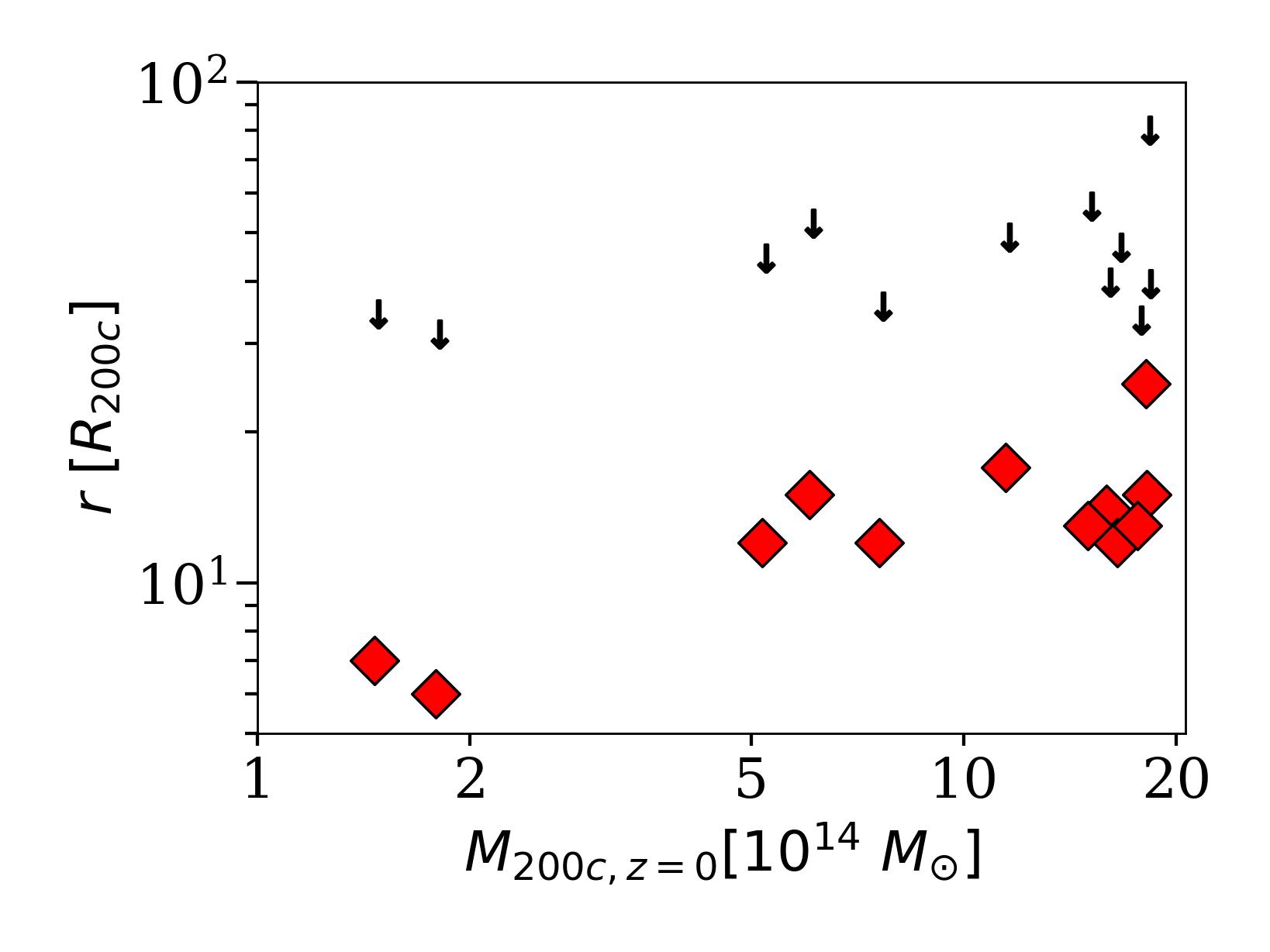}
    \caption{Extension of the PC regions (in units of $ \rm R_{200c}$ of the main progenitor halos) at $z=2.2$ as a function of the mass, $ \rm M_{200c}$, of the descendant clusters at $z=0$ (red diamonds). The downward arrows show, for each progenitor, the boundary of the high-resolution region (see main text).}
    \label{fig:extensionPC}
\end{figure}

\begin{figure*}
 \centering
 \begin{subfigure}[b]{0.49\textwidth}
   \centering
   \includegraphics[width=\textwidth]{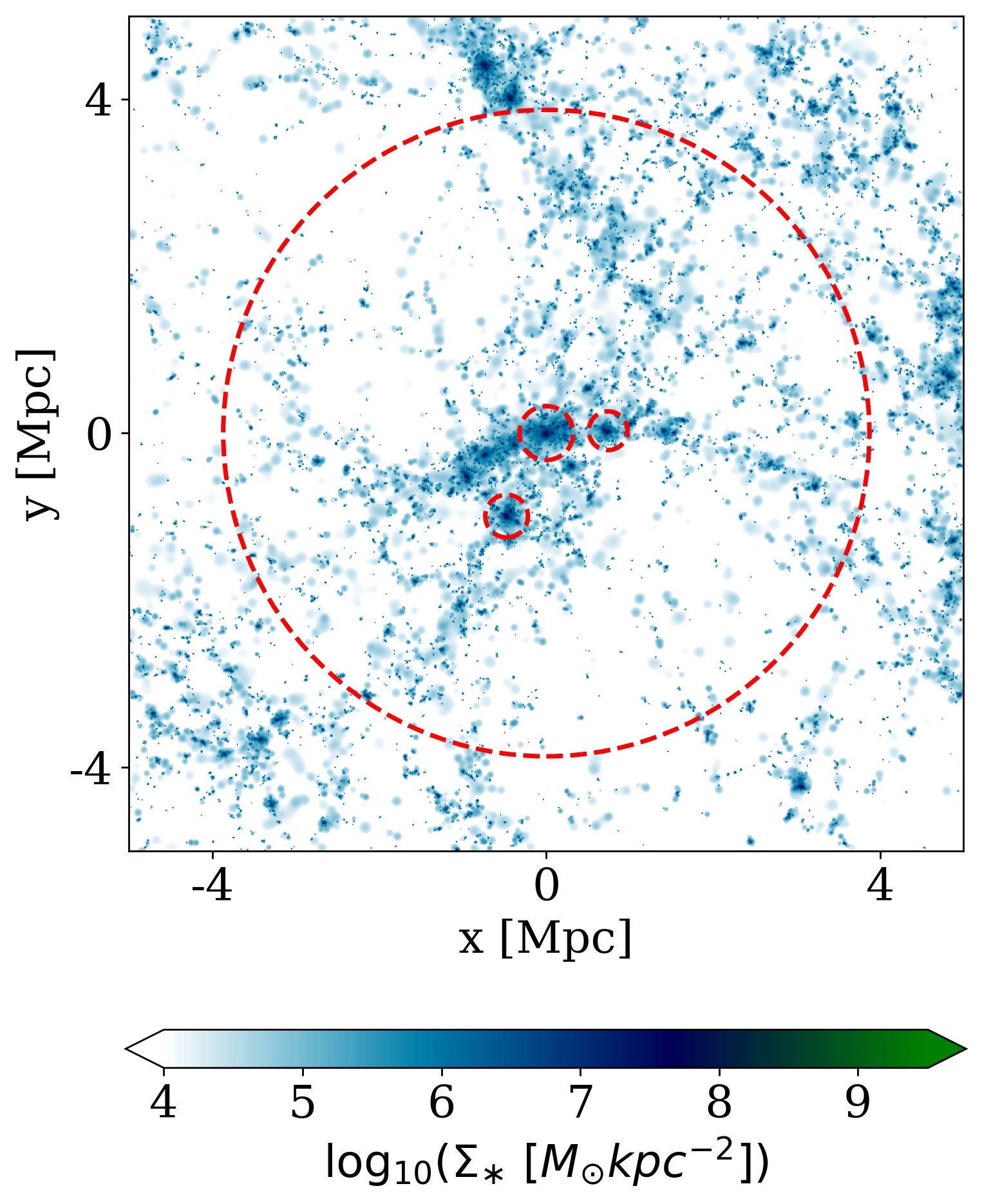}
 \end{subfigure}
 \begin{subfigure}[b]{0.49\textwidth}
  \centering
  \includegraphics[width=\textwidth]{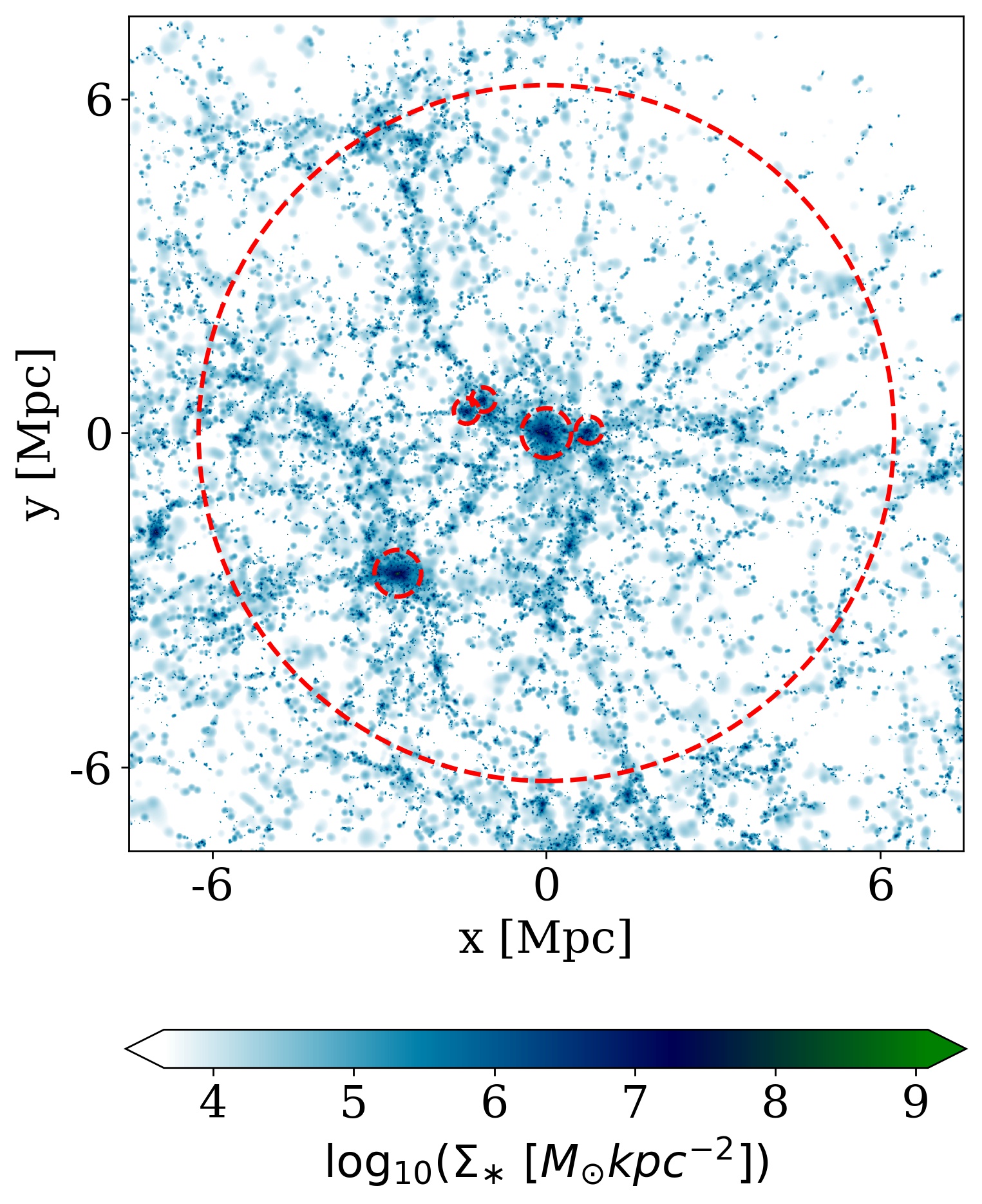}
 \end{subfigure}
 \caption{Projected stellar mass density maps of two {\tt DIANOGA} PC regions, which are the progenitors of clusters of total mass $ \rm M_{200c}$ equal to $ \rm 5 \times 10^{14} \ \text{M}_{\sun}$ and $ \rm 2 \times 10^{15} \ \text{M}_{\sun}$ at $z=0$ (left and right panel, respectively). In each panel, the larger red dashed circle corresponds to the PC region as defined in the text, while the smaller circles show the $ \rm R_{200c}$ of halos in these regions with $ \rm M_{200c}>10^{13} \ M_{\odot}$. Densities are computed in pixels of size 15 (10) kpc in the right (left) panel.}
 \label{fig:stellarmap}
\end{figure*}

This article focuses on characterising cluster progenitors, or protoclusters, at cosmic noon, specifically looking into the snapshots at $z=2.2$ for each simulation. It is worth reminding that no unique and objective definition exists for protoclusters in observations since the fate of observed galaxy overdensities can only be predicted statistically. 
\newline
In the following, we adopt an operational definition of a protocluster (PC) as the volume occupied by the DM that collapses into the final $ \rm R_{200c}$ radius of the $z=0$ cluster.
\newline
To do this, for each cluster we trace back the DM particles that fall within $\rm R_{200c}$ at $z=0$ back to $z=2.2$ and we identify the main progenitor of the cluster as the halo to which the majority of them are associated by the Friends-of-Friends halo finder in {\tt SubFind}. For the two clusters that were not simulated to $z=0$, we assume that the main progenitor at $z=2.2$ is the most massive halo within the high-resolution region. We verified that this is indeed the case for nine out of the twelve regions that were simulated to $z=0$. As for the remaining three regions, in two cases the main progenitor is the second most massive halo (with mass ratio to the most massive of $\sim1.1$), while in the third case it is the fifth most massive halo (mass ratio $\sim 1.8$). 
\newline
After identifying the main progenitor, we define the PC region as the spherical region around the center of the main progenitor halo, encompassing 80$\%$ of the traced DM particles.  This specific value of the percentage of DM particles is clearly arbitrary, but represents a sensible operational definition in order to encompass a region that is large enough to have a complete view of the cluster progenitor environment, while avoiding regions that are dominated to a very large extent by structures that will not be part of the cluster by $z=0$. Indeed, we verified that, when considering the region between the spheres that contain $80$ and $100\%$ of the traced particles, $94\%$, on average, of the particles in this region will not fall within the descendant cluster by $z=0$.  For the two simulations that do not reach $z=0$, we know from the parent DM-only simulation that their masses $\rm M_{200c}$ at $z=0$ are in the range $\rm (2-3) \times 10^{15} \ M_{\odot}$. We then assume that the radial extent of these PC regions is given by the mean radius (expressed in units of $\rm R_{200c}$ of the main progenitor halo at $z=2.2$) of the progenitors of the six most massive clusters, whose masses are in the range $\rm (1-2) \times 10^{15} \ M_{\odot}$. The size of such PCs at $z=2.2$ is approximately $ \rm 15  \times  \ R_{200c}$.  We summarize the properties of the PC regions in Tab. \ref{tablePC}. 
\newline
Figure \ref{fig:dmmap} shows the DM density maps of three {\tt DIANOGA} clusters at $z=0$ (left column), with $z=0$ masses $ \rm M_{200c} = 2\times10^{15} \ M_{\odot}$ (top), $ \rm 9\times10^{14} \ M_{\odot}$ (middle) and $ \rm 2\times10^{14} \ M_{\odot}$ (bottom) and of their $z=2.2$ progenitors (right column). At $z=2.2$, we also show particles ending up within $ \rm R_{200c}$ at $z=0$. The choice of setting the boundary of the PC at the sphere containing 80$\%$ of the total DM particles that are found within $ \rm R_{200c}$ by $z=0$ is sufficient for encompassing all relevant DM structures that are merging into the final cluster. However, the assumption of spherical symmetry is not always optimal, as shown in the top right panel of Fig. \ref{fig:dmmap}. This PC region exhibits two halos of comparable size, and the center of the main halo here does not approximate the center of mass of the distribution of the  traced  DM particles. Nevertheless, we stick with this definition by analogy with PCs defined in observations, often identified around the most massive overdensity observed in a PC region.
\newline
We show the radius of the PC regions as defined above in Fig. \ref{fig:extensionPC}, in units of $\rm R_{200c}$ of the main progenitor halo, as a function of the final ($z=0$) mass of the cluster. We also show in Fig. \ref{fig:extensionPC} the distance from the center of the defined PC region to the closest boundary of the high-resolution region, defined as the distance of the closest low-resolution DM particle from the center of the main progenitor halo. We point out that this is a conservative definition of the boundary of the high-resolution zoom-in region. In fact, few low-resolution DM particles could spuriously be present within the regions where high-resolution particles are largely dominant. We verified that no such particles are present within the defined  PCs, thus guaranteeing that the chaotic diffusion of low-resolution particles does not contaminate these regions. \newline
 As expected, there is a correlation between the extension of the PC radius and the final mass of the cluster it will collapse into, reflecting the hierarchical build-up of structures. On the other hand, the arbitrariness of the PC center for large systems, like the one shown in the top panel of Fig. \ref{fig:dmmap}, produces an artificial increase of the sphere's radius for more massive systems.
\newline
Figure \ref{fig:stellarmap} shows the projected maps of stellar mass densities in the progenitors of two example clusters, with $ \rm M_{200c}=5\times10^{14} \ \text{M}_{\odot}$ and $ \rm 2\times10^{15} \ \text{M}_{\odot}$ at $z=0$. Both regions are characterized by filaments of galaxies intersecting each other at the locations where massive halos are emerging.
\newline
In this work, we analyze the properties of simulated PC galaxies and compare them with analogs at the same redshift in a reference, average density region.
To this purpose, we will use the population of galaxies identified within a cosmological box with size 49 cMpc/h per side, simulated with the same version of {\tt OpenGADGET3}, assuming the same cosmology and subgrid models for baryonic physics. To have the same mass resolution as the {\tt DIANOGA} zoom-in regions, initial conditions have been generated for $576^3$ DM particles and as many gas particles.  In this box, two cluster-size halos of masses $\rm M_{200c}=1.8$ and $1.7 \times 10^{14}$~M$_{\odot}$ are present at $z=0$; however, we did not exclude their progenitors from the analysis of the galaxy population, following the ``field'' definition adopted in observational studies, which refers to a large, representative volume, with all the possible overdense structures it contains. 
\begin{table}[ht]
\centering
\caption{  Properties of the {\tt DIANOGA} protoclusters.}
\begin{tabular}{lcccc}
\label{tablePC}
ID & $\rm M_{200c,z=0}$ & $\rm M_{200c,\text{MP}}$ & radius & $ \rm N_{c}$ \\
 & $[10^{14}$~M$_{\odot}]$ & $[10^{13}$~M$_{\odot}]$ & [Mpc]& \\
 \hline
1 & 5.2 & 3.4 & 3.9 & 2 \\
2 & 7.6 & 4.8 & 4.4 & 3 \\
3 & 6.0 & 1.8 & 4.0 & 1 \\
4 & 1.8 & 4.0 & 2.1 & 1 \\
5 & 16 & 8.9 & 6.2 & 2 \\
6 & 16 & 12 & 6.0 & 4 \\
7 & 1.5 & 2.8 & 2.1 & 1 \\
8 & 15 & 7.0 & 5.3 & 6 \\
9 & 18 & 3.2 & 7.9 & 8 \\
10 & 18 & 11 & 7.1 & 4 \\
11 & 32 & 9.1 & 6.7 & 7 \\
12 & 18 & 8.3 & 6.5 & 5 \\
13 & 18 & 11 & 6.2 & 2 \\
14 & 11 & 3.4 & 5.5 & 3 \\
\end{tabular}
\tablefoot{{Reported above are the ID, the mass of the $z=0$ cluster, the mass of the main progenitor halo at $z=2.2$, the radius of the PC region and the number of PC cores (halos with $\rm M_{200c}>1.5\times 10^{13} \ M_{\odot}$ within the PC region) for each {\tt DIANOGA} PC. The $z=0$ masses of the clusters with ID 10 and 11, that were not simulated to $z=0$, come from the parent DM simulation.}}
\end{table}

\subsection{Photometric properties of simulated galaxies}
\label{sec:skirt}
In this section, we describe the radiation transfer simulations we performed to produce dust-attenuated spectral energy distributions (SEDs) for the simulated PC galaxies. These are used to analyze rest-frame UVJ color-color diagrams, and in particular compare the results of this widely used photometric classification of star-forming and quenched galaxies in observational studies \citep[e.g.,][]{labbe05,williams09} with the classification based on their intrinsic properties in the simulation. 
A proper comparison with the UVJ photometric classification approach requires accounting for the effect of the extinction produced by dust on the UV/optical/NIR emission associated with the stellar populations in the simulated galaxies. We resort to the Monte Carlo radiation transfer code SKIRT-9\footnote{\url{https://skirt.ugent.be/root/_home.html}} \citep{skirt} for this purpose. 
The simulations that we analyze here do not include a self-consistent treatment for the generation and evolution of dust grains. However, this can be implemented and indeed has been done in a lower-resolution version of these simulations \citep[see][for details]{gjergo18}. Therefore, we model the dust distribution through the spatial distribution of metals in the gas component of the simulated galaxies by adopting a fixed dust-to-metal ratio, for which we adopt the value $\rm f_{DTM} = 0.2$ \citep[e.g.,][]{trcka22}. To each SPH gas particle, with mass $ \rm m_{gas}$ and metallicity $Z$ (fraction of mass in metals), is then attributed a dust mass equal to $ \rm m_{dust} = f_{DTM}\, Z\, m_{gas}$. 
\newline
We adopted the THEMIS model \citep{themis} to assign a composition and grain size distribution to dust in the InterStellar Medium (ISM), with associated optical properties that are based on laboratory measurements of materials physically similar to interstellar dust. To discretize the simulation physical domain, in terms of properties of the absorbing medium, we adopted an oct-tree structure \citep[e.g., ][]{saftly13}. In this kind of grid, cubic meshes are recursively split in half their linear size, thus generating eight ``child nodes'' each, until they meet a convergence criterion in the last ``leaves'' of the tree\footnote{ We considered boxes of 64 kpc per side encompassing each galaxy and chose, for the construction of the oct-tree, a maximum fraction of the total dust mass in a leaf equal to $8\times10^{-7}$, and a maximum number of refinement levels equal to 7, corresponding to a minimum size of leaf nodes equal to the gravitational softening of the gas component in the simulations. }. In each cell of the final grid, the properties of the medium (dust) and of the radiation field originating from the sources, represented by star particles of the hydrodynamical simulations, are considered uniform. The stellar emission is modeled by approximating each star particle as a Simple Stellar Population (SSP), consistent with the stellar evolution model included in the {\tt DIANOGA} simulations. For each SSP we adopted a \citet[][]{bc03} SED, which is specified by its age, initial mass and metallicity, as predicted by the simulations, once a \citet{chabrier} IMF is assumed. 
\newline
The self-consistent treatment described above does not account for age-dependent extinction \citep[e.g.,][]{silva98,granato01}, with young stars highly attenuated by the remnants of their high-opacity birth clouds on spatial scales which are not resolved by the cosmological simulations. To gauge the impact of age-dependent dust extinction on our results, we also ran simulations with SKIRT, including a model for molecular clouds (MCs) and the self-consistent treatment of the diffuse ISM described above. We followed \citet{granato21} but adopted a higher age threshold of 10 Myr to select stars still affected by their birth cloud, as in \citet{baes24}. 
To model the extinction of young stars, we placed them all at the centre of an idealized macro-MC with a fixed $ \rm M/r^2$ ratio, where M is the molecular gas mass, which we approximated with the total mass of the ``cold''  gas phase of the ISM (SH03) of each galaxy, and r is its radius. Indeed, it can be shown that assuming all young stars to be embedded in one large MC with a given opacity and extracting their cumulative SED is equivalent to considering them singularly attenuated by MCs of the same opacity and then summing  over their SEDs. To assign an opacity to the macro-MC, we start from \cite{silva98}, which showed that the $ \rm M/r^2$ ratio, and not the single values of M and r, determines the opacity of MCs. Then, we constrained the radius of the idealized MC by adopting the same $ \rm M/r^2$ ratio of a giant MC with a mass of molecular gas equal to $ \rm 10^6 \ M_{\odot}$ and a radius of 15 pc, as done in \cite{granato21}. Then, the dust mass in this cloud is given by $ \rm M_{dust,MC} = f_c \, f_{DTM} \, Z_c \,M_{gas}$, where $ \rm f_c$ is the fraction of cold gas in each galaxy, as predicted by our cosmological hydrodynamical simulations, $ \rm Z_c$ is the metallicity of the cold component of the ISM and $ \rm M_{gas}$ is the total gas mass. For consistency, we removed the cold gas assigned to the MCs from the diffuse medium by multiplying the mass of each gas particle (therefore of each dust particle) by $ \rm 1-f_c$.
SEDs of individual galaxies were then computed by summing the SEDs of the old and young star populations. 
\newline
In both the runs with and without a model for MCs, we ran SKIRT in extinction-only mode, where calculations on the dust thermal re-emission at IR wavelengths are omitted to speed up the computations. We defer to future work a full exploration of the parameter space concerning the sub-resolution modeling of MCs, which is beyond the purpose of this analysis. In Sect. \ref{sec:uvj} we show that the  
 conclusions presented in this work are not affected by the modeling of MCs.

%
\section{Results}
\label{sec:results}
\subsection{Properties of protocluster halos}

In this section, we present the properties of the population of halos in {\tt DIANOGA} PCs, specifically focusing on their masses and characteristic radii. We compare these properties with those of observed halos associated with PCs or high-$z$ clusters, which we will use as a reference for comparing the simulated galaxy population properties in the following sections. Additionally, we examine the velocity dispersion of their member galaxies to provide a more complete comparison with dynamically inferred masses and to identify potential sources of discrepancy in such estimates. Indeed, velocity dispersions are commonly used in dynamical studies of observed halos as proxies for their total masses, assuming spherical symmetry and local relaxation. In the simulated PCs, we consider all halos with $\rm M_{200c}>10^{13} M_{\odot}$, and calculate the 1D velocity dispersions of galaxies identified within the 3D $\rm R_{200c}$ of each halo, along three orthogonal lines of sight. \newline
\begin{figure}
    \centering
    \includegraphics[width=1.1\linewidth]{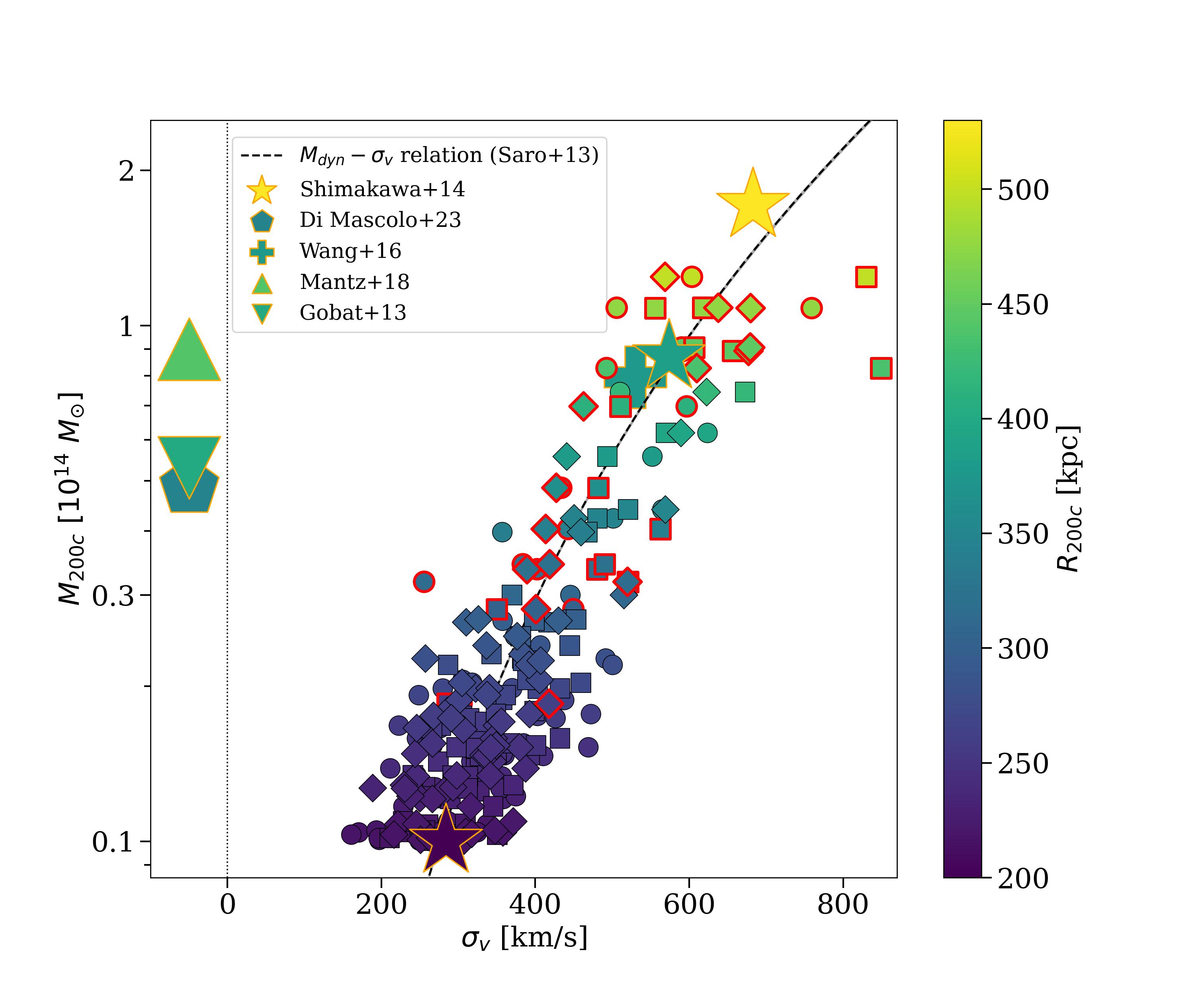}
    \caption{Mass $\rm M_{200c}$ of massive halos in PC regions, as a function of velocity dispersion $\rm \sigma_v$ along three orthogonal lines of sight (represented by circles, squares and diamonds), as traced by galaxies with stellar masses larger than $\rm 10^{9} \ M_{\odot}$ and lying within the $\rm R_{200c}$ of each halo. Each point is color-coded according to the halo $ \rm R_{200c}$. The symbols with red contours mark the central halos of all PCs. The dashed curve represents the relation between dynamical masses and galaxy velocity dispersions calibrated in simulations by \citet{saro13} up to $z=1.2$ and extrapolated at $z=2.2$. Star symbols represent velocity dispersions and dynamical mass estimates within the core of the Spiderweb protocluster and for the two subgroups of USS1558-003 from \citet{shimakawa14}. The cross symbol corresponds to CLJ1001 \citep{wang16}. The pentagon symbol shows $ \rm M_{200c}$ for the Spiderweb PC, from measurements of the ALMA Sunyaev-Zeldovich effect \citep{dimascolo23}, the triangle is for the cluster XLSSC 122 \citep{matz18,willis20,noordeh21} and the downward triangle is for ClJ1449 \citep[][]{gobat13}. These three points are shown artificially at $\rm \sigma_v<0$ since velocity dispersions measured consistently within the estimated radii are not available.}
    \label{fig:dispersions}
\end{figure}
We show the results of this analysis in Fig. \ref{fig:dispersions}, where the masses $\rm M_{200c}$ are plotted as a function of the galaxy velocity dispersions along the three lines of sight. These results are compared with observational estimates for (proto)clusters at $z\sim2-2.5$ from \cite{gobat13}, \cite{shimakawa14}, \cite{wang16,wang18}, \cite{matz18} and \cite{dimascolo23}. The estimates from \cite{shimakawa14} for the Spiderweb protocluster at $z=2.2$ and the two subgroups of USS1558-003 at $z=2.53$, within their estimated $ \rm R_{200c}$, come from a dynamical analysis based on velocity dispersions. The remaining data include estimates based on observations of the Intra-Cluster Medium (ICM): X-ray data for CLJ1001 at $z=2.5$ \citep[][in agreement with estimates based on stellar mass and velocity dispersions]{wang16}, XLSSC 122 at $z=1.98$ \citep{matz18,willis20,noordeh21}, and CLJ1449 at $z=2.00$ \citep{gobat13,valentino16,strazzullo18}, as well as estimates from the Sunyaev–Zeldovich (SZ) signal for the Spiderweb protocluster \citep[][consistent with X-ray measurements from \citealt{tozzi22b}]{dimascolo23}.  Wherever needed, we convert $ \rm M_{500c}$ to $ \rm M_{200c}$ (and $ \rm R_{500c}$ to $ \rm R_{200c}$) using the relations from \cite{ragagnin21}. \newline
We also show an analytic redshift-dependent relation between dynamical mass and velocity dispersion, calibrated by \citet{saro13} on simulations up to $z=1.2$, and extrapolated to $z=2.2$, in Fig. \ref{fig:dispersions}. In general, the simulated halos are in good agreement with this relation. For the few large halos with $\rm M_{200c} \sim 10^{14} \ M_{\odot}$, the figure suggests an increased scatter in velocity dispersions, though with very limited statistics. This might be due to the earlier stage of assembly of these objects. In such cases, the assumption of spherical symmetry underlying the definition of $\rm R_{200c}$ may not be valid. The comparison with \citet{shimakawa14} shows good overlap with the mass and velocity dispersion ranges of the simulated halos studied here, with the Spiderweb PC core, lying at the high-mass end of our probed range, a factor $\lesssim 2$  more massive than our most massive halos. On the other hand, this estimate of the Spiderweb PC core mass is a factor $\sim3$ larger than the ICM-based estimates \citep{tozzi22b,dimascolo23}, which are well within the mass range of our simulated halos, as well as the other halos mentioned above.\newline
The inconsistency between different observations of the same system is likely due to the complications of a dynamical analysis in highly disturbed systems, especially at these redshifts. Such analyses rely on the assumptions of local virialization, spherical symmetry, and the reliable removal of line-of-sight interlopers, all of which may bias the measurement of the velocity dispersion. \newline
With these caveats, we conclude that our PC regions are populated with halos similar in mass and velocity dispersion (when available) to clusters and  protocluster cores identified in observations at $z \sim 2-2.5$.

\subsection{Galaxy Stellar Mass Function}
\label{sec:gsmf}
The Galaxy Stellar Mass Function (GSMF) is an important observable in constraining galaxy formation models since it is essentially the integrated star formation history in a given region. In cosmological simulations, the GSMF at various redshifts is often used to fine-tune parameters of sub-resolution models related to the ISM and star formation processes. However, in the {\tt DIANOGA} simulations,  no tuning has been applied to the evolution of the GSMF, making it a genuine prediction.  \newline In this section, we present the simulated GSMF and investigate its environmental dependence by selecting galaxies from different regions within the PCs and in the cosmological box. We also compare the simulated GSMF with observed GSMFs in PCs at $z \sim 2-2.5$, in order to assess whether our simulations can reproduce the integrated star formation history observed in these systems.
\newline
\begin{figure}
    \includegraphics[width=1\linewidth]{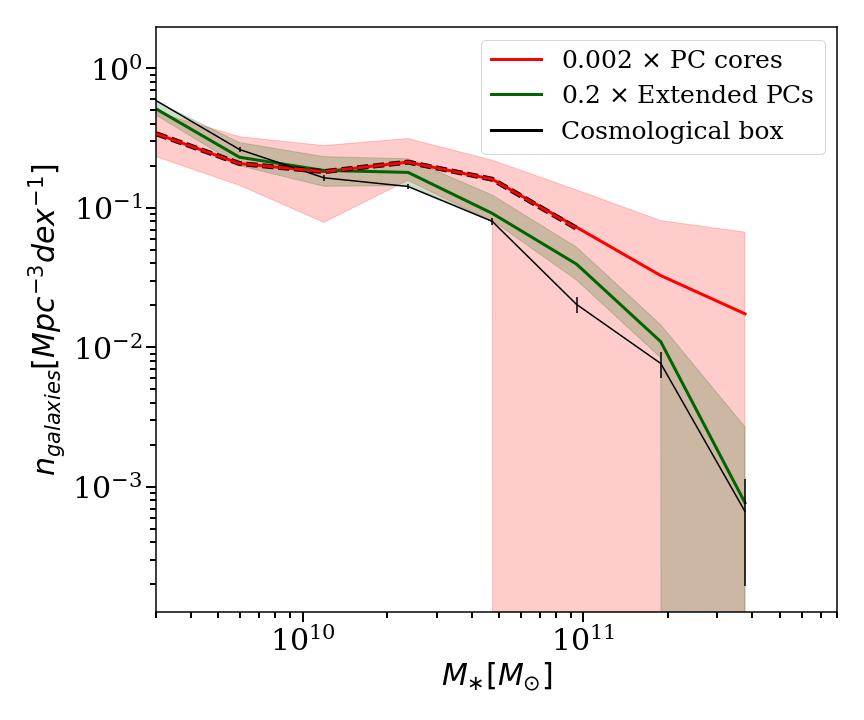}
    \caption{GSMFs in the PC environment, compared with the cosmological box. The red line is the median of the GSMFs in PC cores, with and without central galaxies (solid and dashed lines, respectively; we note that the dashed line stops at $ \rm M_{\ast}\sim 10^{11} M_{\odot}$ because the median GSMF in PC cores in the higher mass bins is zero). The green line shows the median stellar mass function in the less dense, diffuse PC regions, masking out all PC cores.  The normalizations have been rescaled by the factors reported in the legend. The shaded regions define the  16th-84th  percentile range spanned by the different PCs (or PC cores). The black line is the GSMF in the cosmological box, with Poissonian uncertainties.}
    \label{fig:gsmf}
\end{figure}
\begin{figure}
    \centering
    \includegraphics[width=1\linewidth]{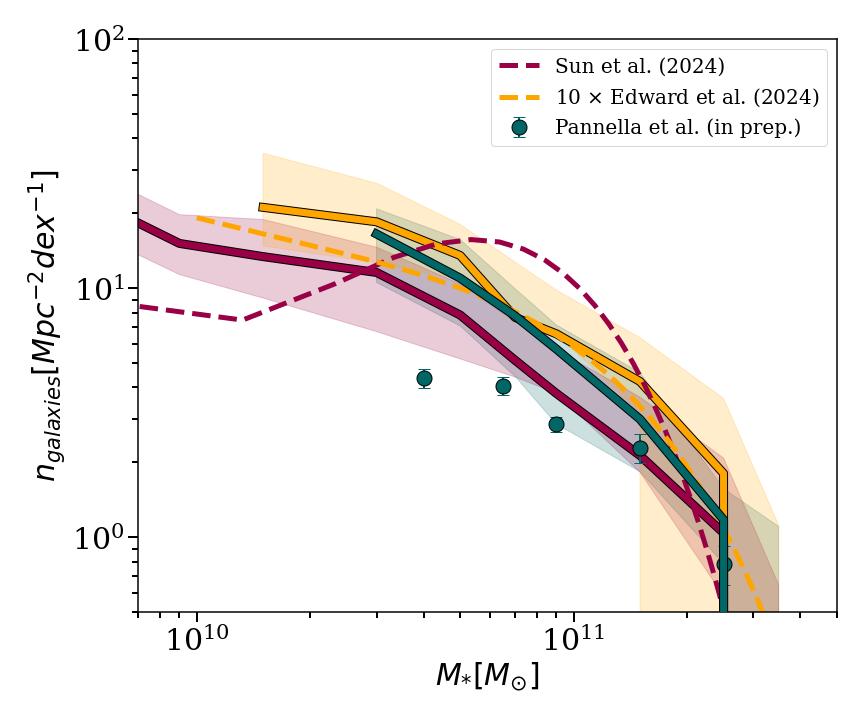}
    \caption{Comparison between the stellar mass function in {\tt DIANOGA} PCs and in observations of PCs at z$\sim$2-2.5 \citep[Spiderweb, CLJ1001, 14 PCs in the COSMOS field; Pannella et al. in prep;][respectively]{sun24,edward24}. For each set of observations, a simulated GSMF has been produced to match the selection of galaxies as closely as possible. For each selection, GSMFs have been extracted along three orthogonal lines of sight, convolved with Gaussian errors with scatter 0.3 dex (see text), and the solid lines (shaded regions) show the median GSMFs ( 16th-84th  percentile ranges, respectively) across all realizations, in the same color as the corresponding data.}
    \label{fig:gsmf-obs}
\end{figure}
The simulated GSMF for {\tt DIANOGA} PCs is shown in Fig. \ref{fig:gsmf}. To highlight environmental signatures associated with the most massive halos (cores) in the PC regions, we separately consider galaxies hosted by halos with $ \rm M_{200c}>1.5\times10^{13} \ M_{\odot}$ in PCs.   
This separation is intended to investigate differences between galaxy populations in PC cores and those in the extended PC structure. The specific threshold of $ \rm M_{200c}=1.5\times10^{13} \ M_{\odot}$ corresponds to the mean mass of the third most massive progenitor halo, with the median mass ratio between first and third most massive progenitors at this redshift being 3.5. We show the predicted GSMF for PC cores, for the extended PC regions\footnote{To calculate the GSMFs of PC cores we used the volume of the sphere with radius $ \rm R_{200c}$ of each halo, while for the extended PCs we used the radius of the PC region as defined in Sect. \ref{sec:pcid}, to which we subtract the cumulative volume occupied by the PC cores.} and for the cosmological box. The GSMF normalization is obviously higher in denser regions; the GSMFs shown in Fig. \ref{fig:gsmf} have been rescaled to highlight differences in their shape.
\newline
The comparison between the GSMFs in the cosmological box and in the extended PC regions reveals a slight relative excess of massive galaxies, with $\rm M_{\ast}>10^{10} \ M_{\odot}$, in the latter. This indicates that even when excluding galaxies in large halos, PCs exhibit a faint signature of accelerated evolution at earlier times. As expected, the excess of massive galaxies is more pronounced for the median GSMF in PC cores.  This signal is completely dominated by central galaxies at $ \rm M_{\ast} \gtrsim 10^{11} \ M_{\odot}$. 
This result is qualitatively consistent with recent observational findings, such as the top-heavy feature in the GSMF of CLJ1001 at $z=2.5$ \citep{sun24} and the large ratio of high-mass to low-mass galaxies relative to the field in the Elentári proto-supercluster at $z \sim 3.3$ \citep{forrest24}. We note that the GSMF of PC cores shows significant variability at high stellar masses among different cores, with a considerable scatter towards low number densities at $ \rm M_{\ast} \gtrsim 5 \times 10^{10} \ M_{\odot}$, which seems to be primarily due to the lower-mass PC cores not having formed a significant fraction of massive galaxies yet.  We note that the presence of massive halos in the cosmological box does not impact the global properties of the galaxy population. We identify the three most massive halos, with $ \rm M_{200c}$ in the range $ (1 - 3) \times 10^{13}$~M$_{\odot}$, in the cosmological box and define a protocluster region around each of them, assuming an extent of 6.5 times their $ \rm R_{200c}$ (the average radius of the {\tt DIANOGA} PCs with similar descendant mass). The galaxies within such regions contribute only to $4\%$ of the total galaxy population within the box, and their presence has only a negligible effect on the GSMF. 
\newline
In Fig. \ref{fig:gsmf-obs} we compare the predicted GSFM with observations in protoclusters at similar redshift, including the Spiderweb protocluster (Pannella et al., in prep.), CLJ1001 \citep{sun24} and 14 protoclusters identified in the COSMOS field at $2<z<2.5$ \citep{edward24}. We try to replicate the selection of the observational data as closely as possible for an accurate comparison between simulations and observations. For the comparison with Pannella et al. (in prep.) and Sun et al. (2024), we compute the GSMF within the same projected area covered by the observations and apply a redshift cut (including the contribution from peculiar motions) equivalent to the one imposed by the narrow-band filters in the data.
\cite{edward24} use photometric redshifts with statistical subtraction of the background, meaning that the actual extension of the probed PC regions along the line of sight is unconstrained. As a result, for comparison to this work we only match the projected area of these observations and rescale their GSMF fit by a factor of 10 to approximately match the normalization of the simulated GSMF. For each simulation, we repeat the selection along three orthogonal lines of sight, extracting the median across all $3\times14$ realizations of the GSMF and compute the  16th-84th  percentiles to define the scatter among simulated PCs. We apply a convolution with a Gaussian distribution of errors in the logarithm of stellar masses, with a dispersion of 0.3 dex, to approximate observational uncertainties (e.g., \citealt{conroy13}, \citealt{mobasher15}, \citealt{lower20}).
For \cite{sun24}, we show the sum of the observed GSMFs of star-forming and quenched galaxies, although the latter is affected by incompleteness. \newline
This comparison highlights an overall agreement in the normalization of the GSMF with observations \citep[excluding][where the normalization has been arbitrarily rescaled as mentioned above]{edward24}. The shape of the simulated GSMF is in very good agreement with \cite{edward24} and consistent with Pannella et al. (in prep.) especially at high masses \cite[note that these observations may suffer from significant incompleteness at lower stellar masses; see][]{tozzi22a}. On the other hand, we do not reproduce the strong top-heavy feature observed in the GSMF in CLJ1001 \citep{sun24}.

\subsection{Star-forming Main Sequence}
\label{sec:ms}
\begin{figure}
    \centering
    \includegraphics[width=1\linewidth]{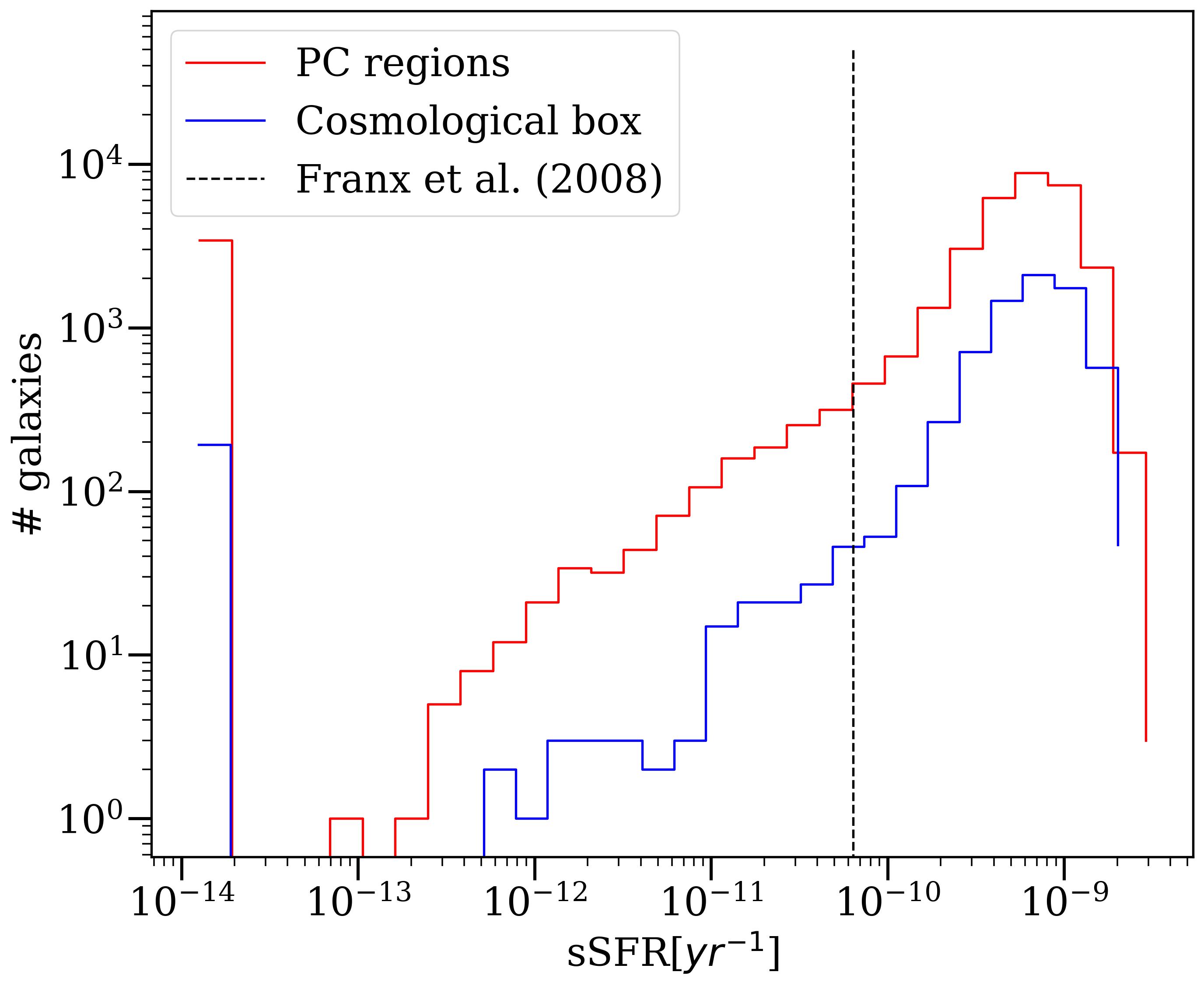}
    \caption{Distributions of sSFRs of galaxies in the {\tt DIANOGA} PC regions (red line) and in the cosmological box (blue line) at $z=2.2$, compared with the threshold separating star-forming from quenched galaxies, at this redshift, from \cite{franx08}. Galaxies with SFR=0 have been artificially assigned the minimum non-zero sSFR, corresponding to $ \rm 10^{-14} \ yr^{-1}$.}
    \label{fig:ssfr_hist}
\end{figure}

\begin{figure*}
 \centering
  \begin{subfigure}[b]{8.5cm}
  \includegraphics[width=8.5cm]{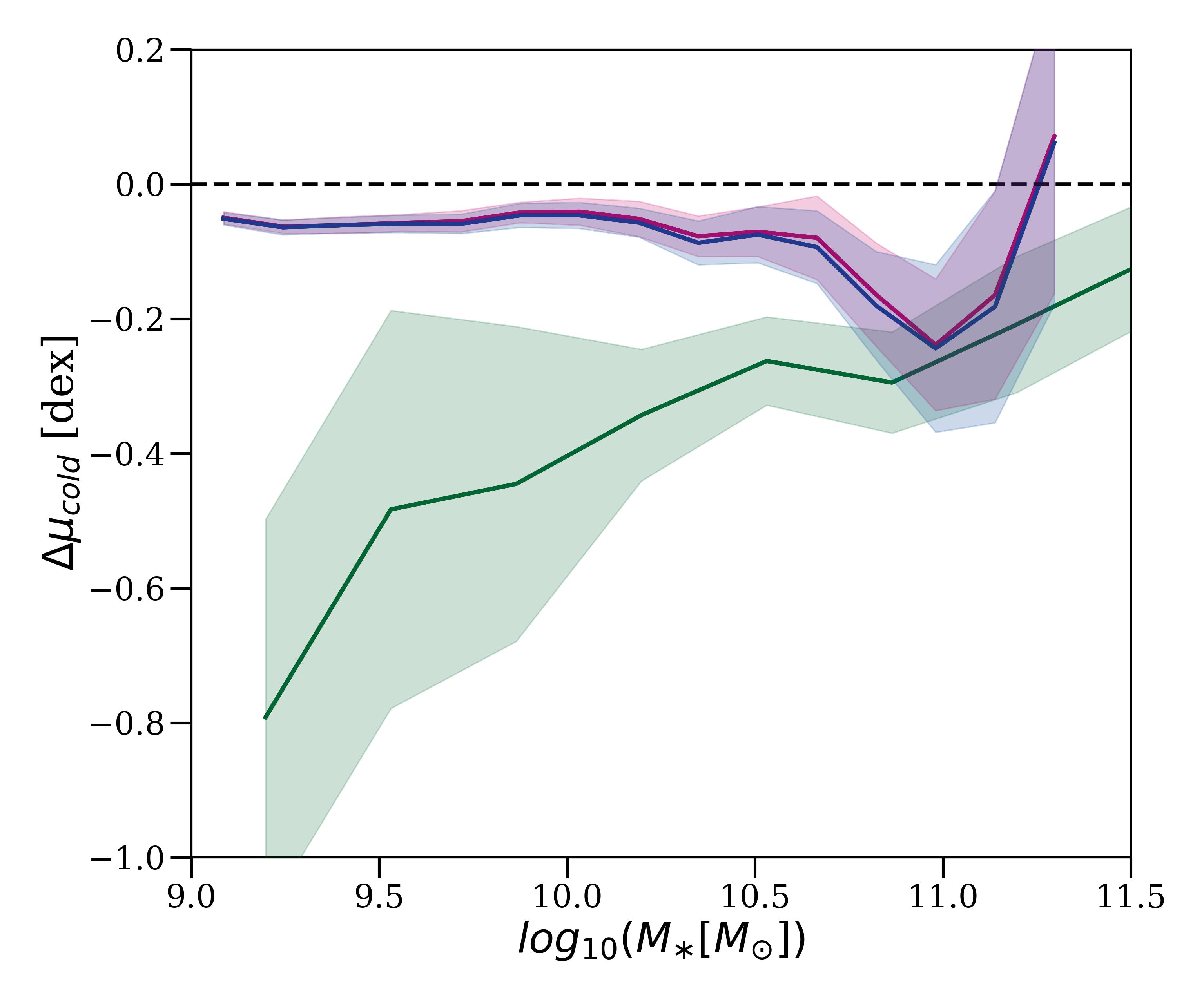}
 \end{subfigure}
 \begin{subfigure}[b]{8.5cm}
   \includegraphics[width=8.5cm]{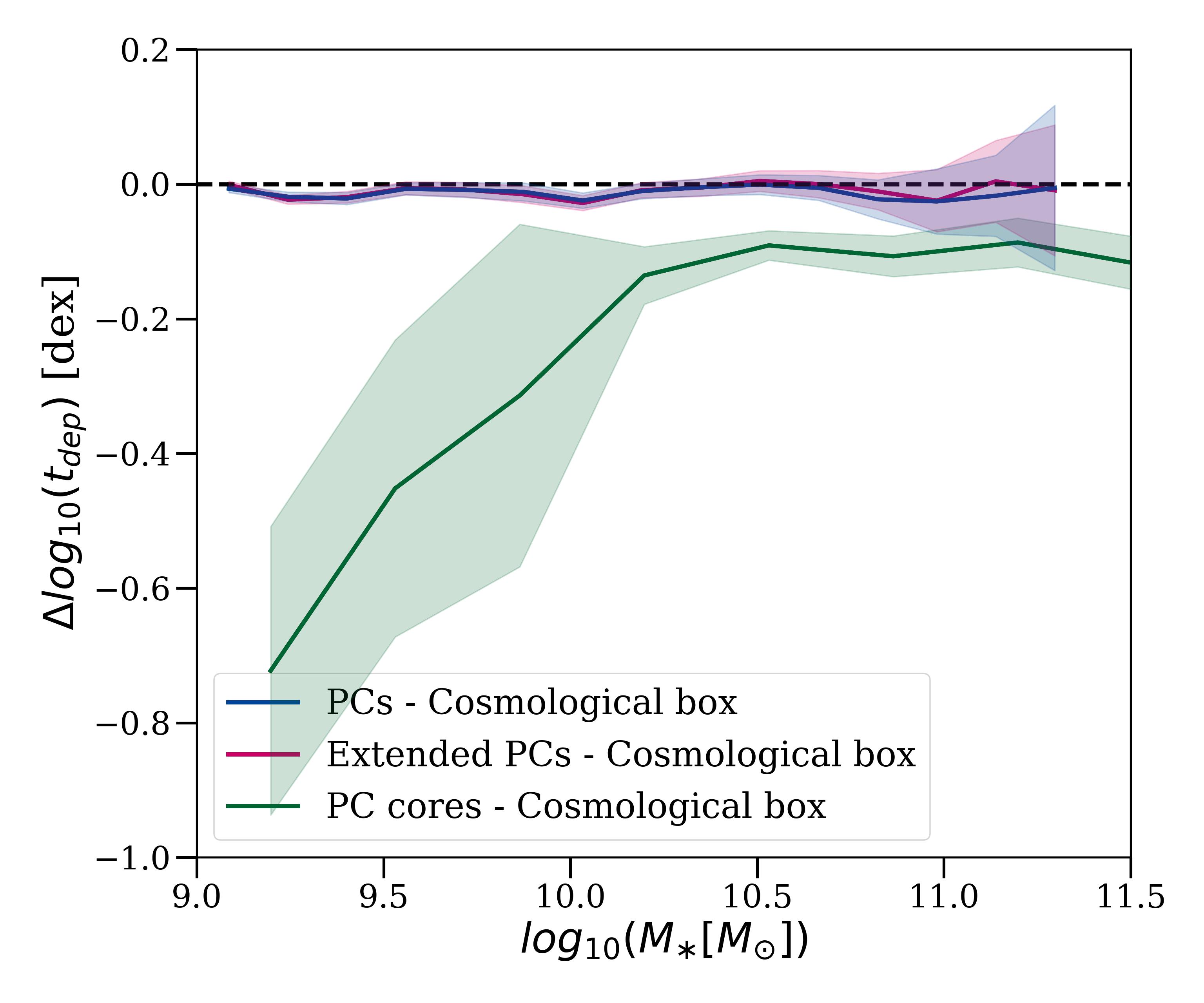}   
 \end{subfigure}
 \caption{Environmental dependence of the fraction of cold gas (left panel) and of the cold gas depletion time (left panel) at  $z=2.2$, as a function of stellar mass. Here we consider four environments: the cosmological box, the whole PCs, the low-density extended PCs and the PC cores. The solid lines define the offsets between the cold fractions and the depletion times in the PC environments with respect to the cosmological box, as defined in the legend, resulting from 1000 Monte Carlo realizations (see text). The shaded regions represent the intrinsic scatter on the offsets.}
 \label{fig:sfr_drivers_offsets}
\end{figure*}

The star-forming Main Sequence (MS) is a key observable for constraining the mode of star formation as a function of stellar mass at different epochs. As such, it is useful to highlight the effect of environment on the star formation activity of galaxies. However, theoretical models continue to struggle to match the normalization of the observed MS at cosmic noon, with most semi-analytic models and cosmological simulation suites underestimating galaxy SFRs at high redshift \citep[see, e.g.,][]{granato15,hirschmann16,donnari19,akins22,andrews24,fontanot24}, despite the fact that simulations can be tuned to reproduce the observed GSMFs quite well.
Star formation has been investigated in lower-resolution versions of the {\tt DIANOGA} simulations, showing indeed lower SFRs of galaxies in PCs with respect to the observed MS \citep[e.g.,][]{bassini20}, and a star formation history not sufficiently bursty, and thus unable to reproduce FIR observations in high-$z$ PCs \citep{granato15}.
\newline
In the following, we compare the star-forming MS in the {\tt DIANOGA} PCs and in the cosmological box at $z=2.2$ to investigate the impact of environment on the population of star-forming galaxies, and compare these results with observations both in the field and in PCs.
We classify galaxies as star-forming if their sSFR exceeds $ \rm 0.3/t_H$ \citep[where $ \rm t_H$ is the Hubble time at $z=2.2$;][]{franx08}. Figure \ref{fig:ssfr_hist} shows that this threshold appropriately separates star-forming and quenched galaxies in our simulations, falling approximately 1 dex below the peak of the simulated MS and intersecting the sSFR distribution at the point where quenched galaxies lead to a change in the slope of the distribution, both in the PCs and the cosmological box.
\begin{figure*}
\centering
\includegraphics[width=1\linewidth]{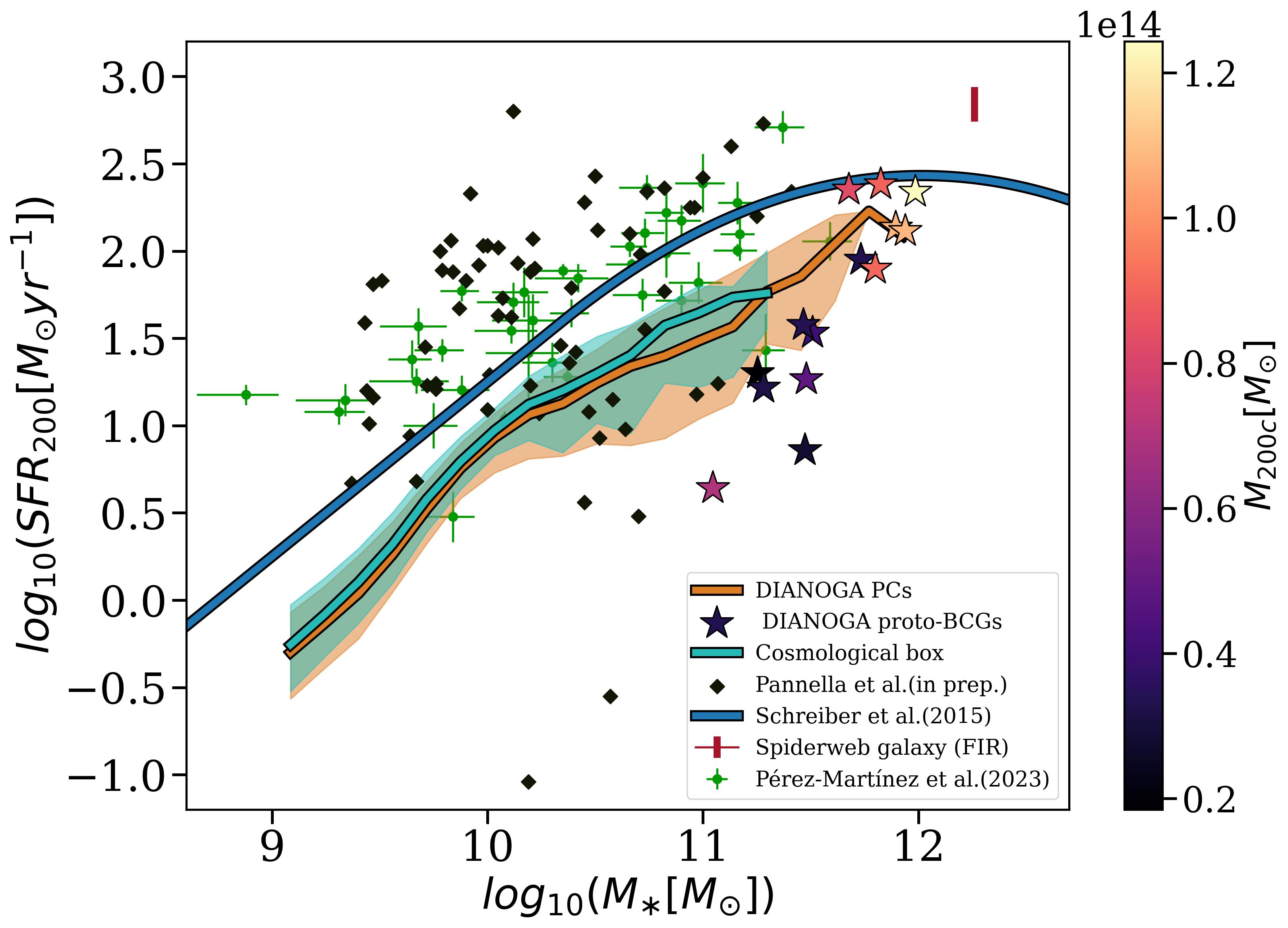}
\caption{The star-forming MS of galaxies in the {\tt DIANOGA} protoclusters (orange) and in the cosmological box (green). Solid lines define the median SFRs of all star-forming galaxies within each stellar mass bin, while the shaded areas mark the 16th-84th percentile interval. The star symbols represent proto-BCGs identified in the simulations, color-coded according to the mass $\rm M_{200c}$ of the main progenitor halo of each protocluster. Black diamonds and green dots are data for the Spiderweb PC from Pannella et al. (in prep.) and from \cite{perezmartinez23}, respectively. The blue line is the best fit for the observed field star-forming MS by \cite{schreiber15} at $z=2.2$. The red bar shows the estimated SFRs for the Spiderweb galaxy from measurements by \cite{seymour12} and \cite{drouart14}. Simulated SFRs are averaged over 200 Myr (see text).}
\label{fig:ms}
\end{figure*}
\begin{figure}
    \centering
    \includegraphics[width=0.95\linewidth]{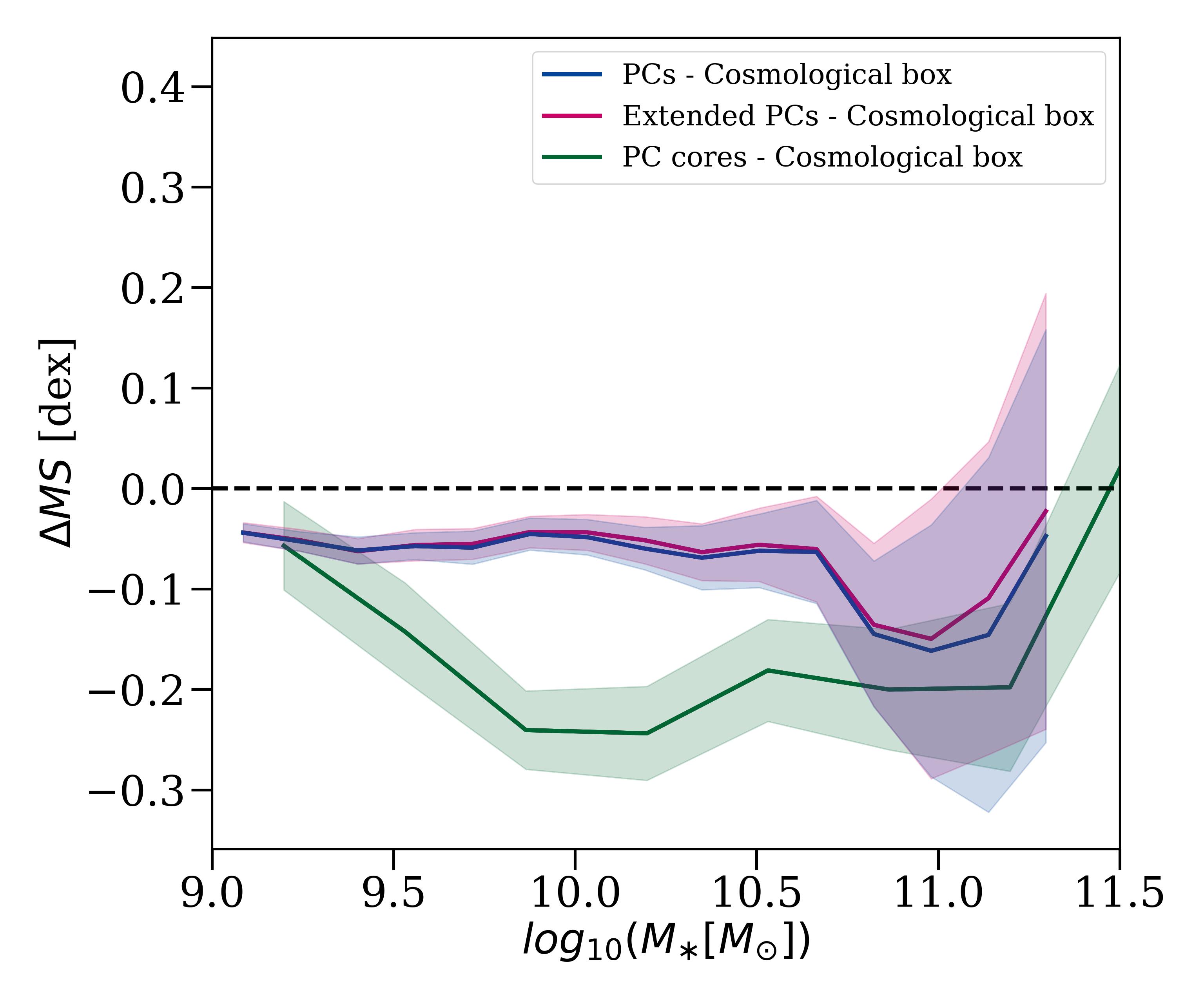}
    \caption{Environmental dependence of the simulated star-forming MS at  $z= 2.2$, as a function of stellar mass. Results are shown for the whole sample of PC galaxies, as well as separating PC galaxies located within/outside $ \rm R_{200c}$ of halos with $ \rm M_{200c}>1.5\times10^{13} \ M_{\odot}$ (PC cores), comparing them to results for the cosmological box. The solid lines and shaded regions show the MS offset and its uncertainty between pairs of environments as indicated in the legend (see text for details).}
    \label{fig:ms_offset}
\end{figure}
\newline
In Fig. \ref{fig:ms} we compare the simulated MS with observational data for protoclusters and the field at the same redshift. 
Specifically, we compare the median SFRs, as a function of stellar mass,  across all the simulated PCs and in the cosmological box   with observations in the Spiderweb PC from Pannella et al. (in prep., with SFR measurements based on dust-corrected rest-frame UV fluxes) and \citet[][with H$\alpha$-based SFRs]{perezmartinez23}, and with field measurements from \citet[][ based on Herschel observations]{schreiber15}. To ensure a more proper comparison with these observations, we compute the SFRs of simulated galaxies averaged over 200 Myr, which is approximately the timescale of the SFR tracers used in Pannella et al. (in prep.) and in \cite{schreiber15}, while H$\alpha$ observations used in \cite{perezmartinez23} trace an almost instantaneous SFR, with the relevant timescales being 5-10 Myr.
We show in Fig. \ref{fig:cfr_sfrs} that the SFRs in the cosmological box and in the PCs at $z=2.2$ decline over the past 200 Myr in our simulations, with instantaneous SFRs systematically lower than averaged ones across all probed stellar masses. Nevertheless, the offset between these two SFRs estimates ($\sim0.1$ dex on average) is much smaller than typical observational uncertainties, therefore,  considering also the limited sample sizes in MS investigations in observed high-$z$ PCs, we expect the comparison between simulated and observed SFRs in Fig. \ref{fig:ms} not to be affected by the different timescales of the mentioned SFR tracers, and thus we use SFRs averaged over 200 Myr in Fig. \ref{fig:ms} and in the following.
\newline
The simulated MS is systematically lower than the observed one, both in the field and in the Spiderweb PC. The 11 star-forming\footnote{The remaining 3 proto-BCGs, also shown in Fig. \ref{fig:ms}, are quenched, according to our definition.} proto-BCGs also have lower SFRs compared to the Spiderweb galaxy. 
A similar underestimation of SFRs at high redshift is present also in lower-resolution simulations of the {\tt DIANOGA} PCs \citep{granato15,ragone18,bassini20}, which also adopt different prescriptions for the feedback processes. This suggests that the primary cause for the lower MS normalization predicted by simulations lies in the star formation model (SH03), which uses a quiescent mode of star formation that inherently prevents any bursty star formation episodes, that are instead expected at cosmic noon. 
\newline
\cite{ragone24} recently showed that by adopting a different model for star formation and stellar feedback \citep[][]{murante10,murante15,valentini20,valentini23}, coupled with $ \rm H_2$ formation on dust grains \citep{granato21}, the discrepancy between the simulated and observed MS can be reduced, also resulting in a cosmic star formation history that better matches the observed one. 
\newline
 Within the simulation, we observe that less massive PC main progenitors host proto-BCGs with suppressed star formation compared to the (simulated) MS. The simulated MS in Fig. \ref{fig:ms} also shows a rather minor suppression of star formation in PCs compared to the cosmological box\footnote{  We verified that masking the ``protocluster'' regions in the cosmological box, as defined above, does not impact this result.}. This is further explored in Fig. \ref{fig:ms_offset} where, for a better understanding of the effect of the environment, we separate star-forming  galaxies hosted by PC cores ($\rm M_{200c} \geq 1.5 \times 10^{13} \ M_{\odot}$) from those in the lower-density extended regions of PCs, comparing them also with galaxies in the cosmological box. 
\newline
 Figure \ref{fig:ms_offset} shows the environmental dependence of the star-forming MS by comparing offsets in SFRs between the aforementioned galaxy populations. To compute the offsets, we generate 1000 realizations of the MS of each galaxy population, assuming an (asymmetric) Gaussian distribution of SFRs as a function of stellar mass based on the median SFR and  16th-84th  percentile range for the given population of simulated galaxies. For each population (and thus environment) pair, we then compute the mean of the SFR offsets as a function of stellar mass over these 1000 realizations. The error on the offset is estimated as the scatter of the SFR offsets across the realizations.
\newline
We predict a very small though significant suppression of approximately 0.05 dex in the SFRs of PC galaxies compared to the cosmological box (we note that this is insensitive to including or excluding the most massive halos, as their galaxy populations are subdominant compared to the full PC structure). This suppression is independent of stellar mass, though for $ \rm log_{10}(M_{\ast}/M_{\odot})>11$ statistical uncertainties are too large to draw any conclusions. Observational uncertainties in current MS determinations do not allow to detect a 0.05 dex offset. Therefore, this prediction from simulations is compatible with studies that found either a low-significance or no suppression in the SFRs of PC galaxies compared to their field counterparts \citep[e.g.,][]{cucciati14,koyama21,polletta21,shi21,perezmartinez23}. On the other hand, our predicted offset contrasts with other studies finding an enhanced star formation in PC galaxies \citep[e.g., Pannella et al. in prep.;][]{shimakawa18a,monson21, perezmartinez24}. 
\newline
Figure \ref{fig:ms_offset} further shows that the star formation suppression is more significant in PC cores, with an MS offset up to $\sim0.25$ dex compared to the cosmological box. This offset decreases to $\sim0.05$ dex for galaxies with $ \rm M_{\ast}\sim10^9 \ M_{\odot}$. For these galaxies, whose star formation history has experienced a more marked recent decline (see Fig. \ref{fig:cfr_sfrs}), the SFR averaged over 200 Myr may be less sensitive to their relatively rapid evolution, especially in PC cores, as will be discussed in the following. 
\newline
Given that star formation in the {\tt DIANOGA} simulations depends on the gas reservoir and on the efficiency of gas conversion into stars, we further explore how these quantities compare with observational constraints. The reservoir of star-forming gas is represented by the cold component of the multi-phase gas particles that describe the ISM in the SH03 star formation model adopted in our simulations (see Sect. 2.1). We compute the cold gas fraction, $ \rm \mu_{cold} = M_{cold}/M_{\ast}$, where $ \rm M_{cold}$ is the total cold gas mass in a galaxy. We compare this cold gas fraction with the molecular gas fraction, $ \rm M_{molecular}/M_{\ast}$, which is estimated in observational investigations, noting that $ \rm M_{cold}$ represents an upper limit to the (sub-resolution) molecular gas content. We also compute the depletion time $ \rm t_{dep} = M_{cold}$/SFR, using the SFR averaged over 200 Myr. 
\begin{figure*}
 \centering
  \begin{subfigure}[b]{8.5cm}
  \includegraphics[width=8.5cm]{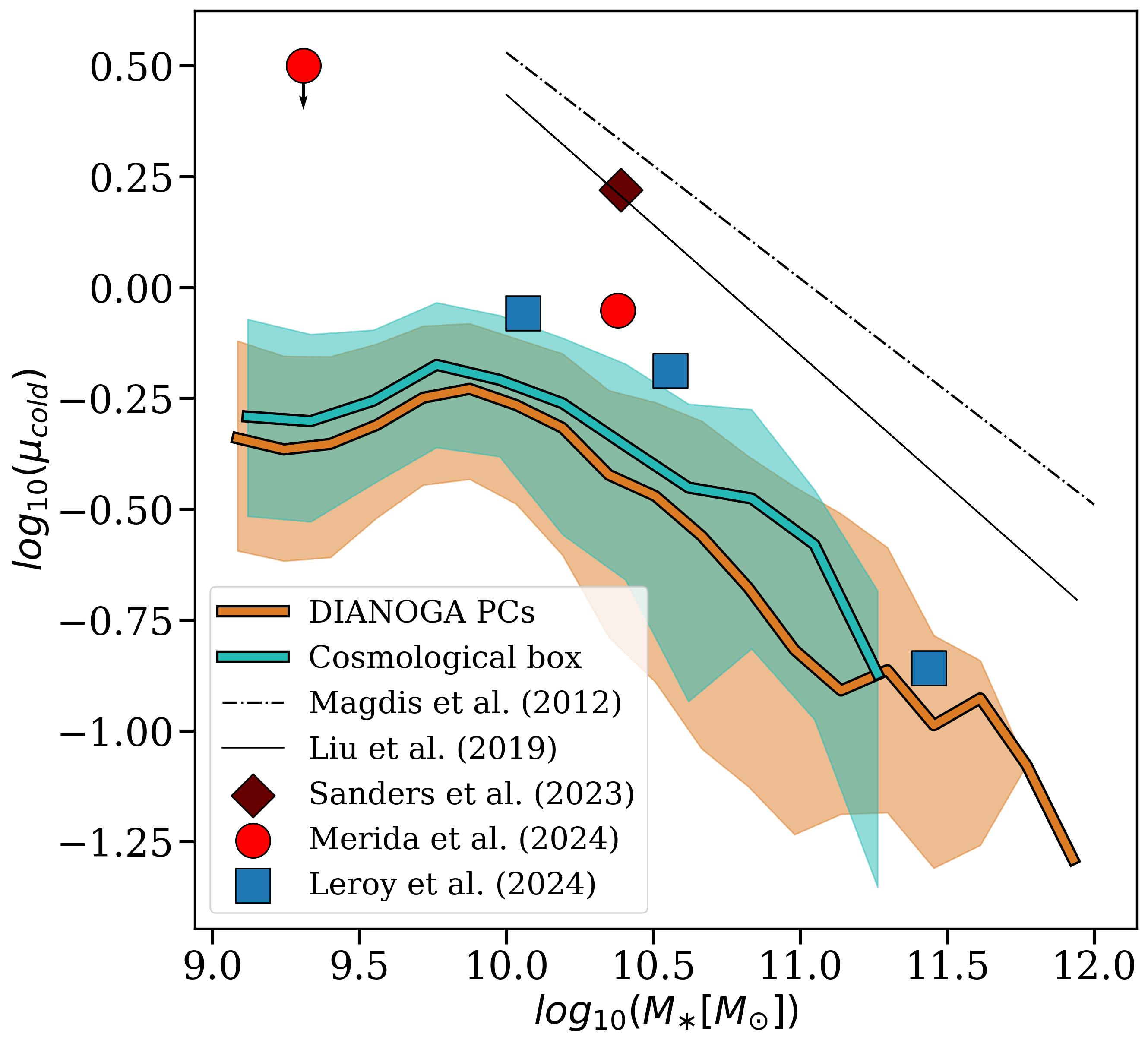}
 \end{subfigure}
 \begin{subfigure}[b]{8.5cm}
   \includegraphics[width=8.5cm]{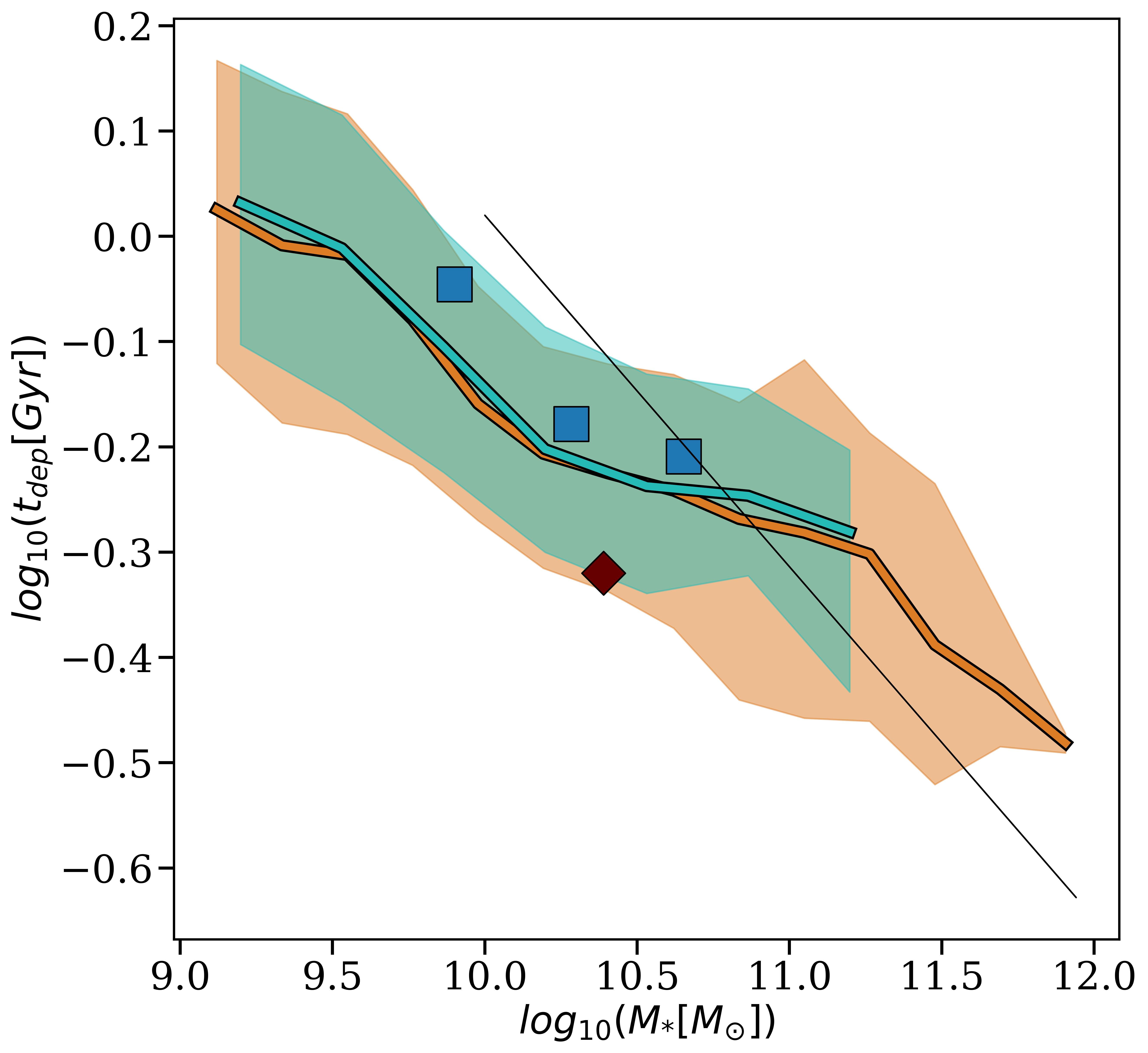}   
 \end{subfigure}
 \caption{Cold gas fractions and depletion times in star-forming galaxies in the {\tt DIANOGA} PCs (orange) and in the cosmological box (green), as a function of stellar masses. Solid lines (shaded regions) show median values ( 16th-84th  percentile ranges, respectively). For comparison, we show observations at $z\sim1.5-3$ from \cite{magdis12}, \cite{liu19}, \cite{sanders23}, \cite{merida24} and \cite{leroy24}.}
 \label{fig:sfr_drivers}
\end{figure*}
\newline
In Fig. \ref{fig:sfr_drivers} we show $ \rm \mu_{\rm cold}$ and $ \rm t_{\rm dep}$ as a function of stellar mass for star-forming galaxies,  both in PCs and in the cosmological box, comparing with observational results from \cite{magdis12}, \cite{liu19}, \cite{sanders23}, \cite{merida24} and \cite{leroy24} at $z\sim1.5-3$. In both PCs and the cosmological box, our simulations predict cold gas fractions that are systematically lower than the observed values, while the depletion times (thus the star formation efficiencies ) appear consistent with observations. However, considering the realistic GSMF in the simulated PCs discussed in Section \ref{sec:gsmf}, the cold gas deficit suggests a possible inefficiency in gas cooling, excessive heating due to feedback processes, or highly efficient interactions with the hot ambient gas, which begins to permeate the simulated halos at these redshifts (Esposito et al., in prep.), rather than an early over-consumption of gas.
\newline
As shown in Fig. \ref{fig:sfr_drivers}, the fraction of cold gas is slightly lower in PCs than in the cosmological box. To investigate this further, we derive offsets in the median cold gas fraction and depletion times of galaxies in PCs in comparison with galaxies in the cosmological box, as a function of stellar mass, using the same methodology applied for the MS offsets in Fig. \ref{fig:ms_offset}. The resulting offsets are shown in Fig. \ref{fig:sfr_drivers_offsets}. For the cold gas fraction, we observe a small ($\sim 0.05$ dex) though significant suppression in the cold gas reservoir in PC galaxies across all stellar masses, similar to the trends seen for the MS. For star-forming galaxies in PC cores, the offset in $ \rm \mu_{cold}$ relative to the cosmological box decreases at lower masses, suggesting a more efficient depletion or less efficient replenishment of gas in low-mass galaxies. Since this feature is specific to the most massive halos in the PC regions, it is possible that interactions with the hot ICM emerging in these massive halos are either stripping cold gas from galaxies or halting cosmological gas accretion, leading to more rapid gas depletion, especially in lower-mass galaxies. In particular, the offset in $ \rm \mu_{cold}$ shows a sharp drop at $\rm log(M_{\ast}) \lesssim 9.5$. This is because approximately 40$\%$ of galaxies in this mass range within PC cores, classified as star-forming based on their averaged SFRs, have completely lost their cold gas reservoir, effectively shutting off star formation over the past 200 Myr. While these galaxies are in a regime where the results could be influenced by the finite numerical resolution of the simulations, galaxies of similar mass in the cosmological box and in smaller halos within the PC regions do not show such an abrupt decrease in cold gas fractions. Therefore, we conclude that this is not a numerical artifact, although the extreme environment in the largest halos within the simulated PC regions, coupled with the relatively low number of gas particles in galaxies with $\rm log(M_{\ast}) < 9.5$, may amplify a genuine physical process occurring in massive halos at this redshift. 
Figure \ref{fig:sfr_drivers_offsets} also shows that the environmental dependence of depletion times qualitatively mirrors the trends observed for the cold gas fraction. The offset in the median depletion time of galaxies in the full PC region in comparison with the cosmological box (with no sensitivity to the inclusion or exclusion of the most massive halos) is marginal and statistically insignificant. However, low-mass PC galaxies within the most massive halos exhibit a dramatic decrease in their median depletion time, corresponding to the low median cold gas fractions discussed above. 

\subsection{Quenched galaxy fractions}
\label{sec:qf}
In this section, we analyze the population of quenched galaxies in the {\tt DIANOGA} PCs and in the cosmological box at $z=2.2$, to quantify how the environment of massive halos in PCs suppresses star formation in galaxies. This represents the first necessary step to exploit the simulations in probing the nature of early quenching processes in cluster progenitor environments, in the context of understanding the physical mechanisms resulting in the predominance of quenched galaxies in mature clusters at later epochs.
\begin{figure}
    \centering
    \includegraphics[width=1\linewidth]{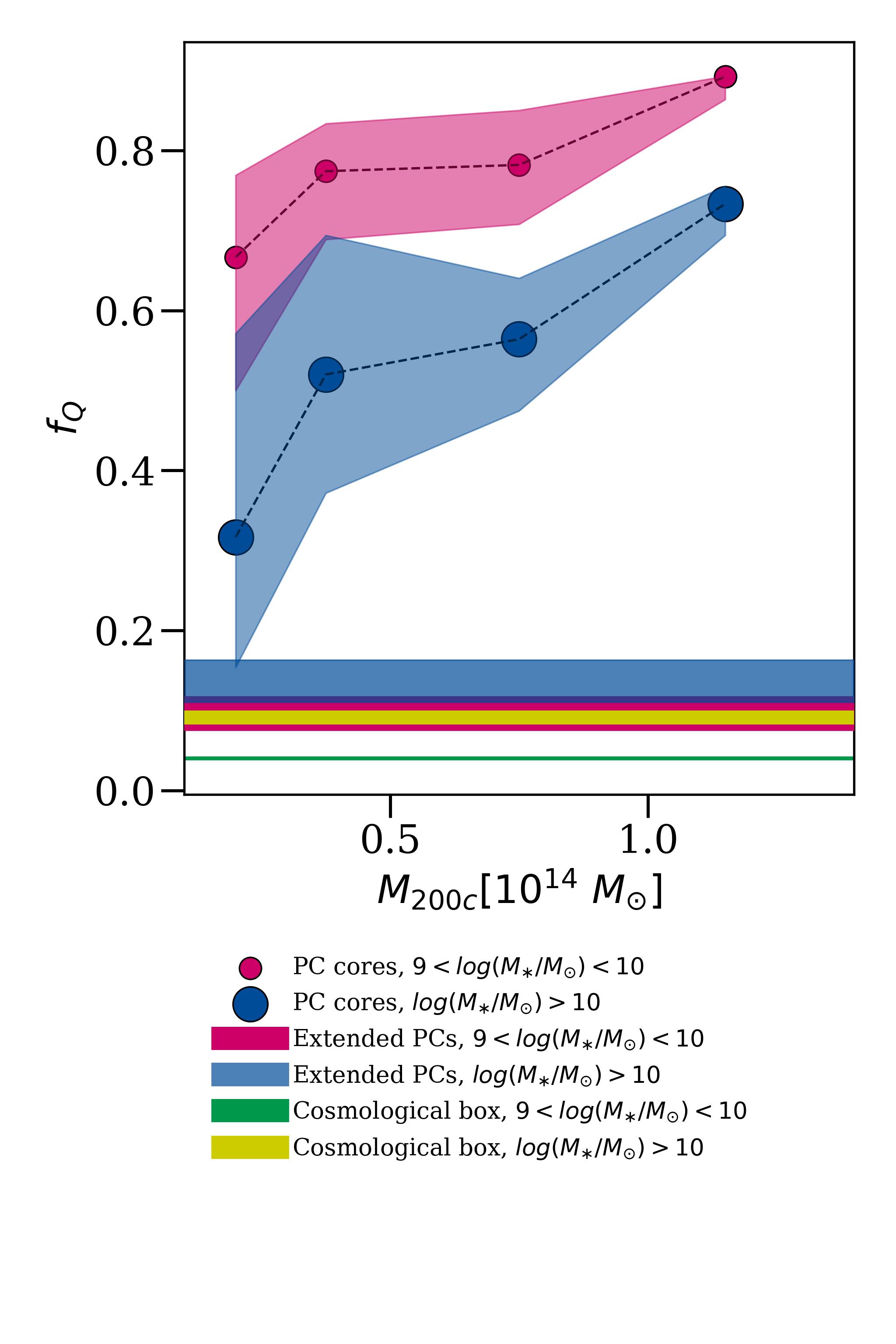}
    \caption{Fraction of quenched galaxies in PC cores, as a function of the core masses $ \rm M_{200c}$, in the {\tt DIANOGA} PCs at $z=2.2$. The dots are median quenched fractions across different cores in each mass bin and the shaded areas are the 16th-84th percentile ranges. The horizontal bands show the fraction of all quenched galaxies in the coeval cosmological box \citep[its width is the $1 \sigma$ confidence interval calculated following][]{cameron11}, and the  16th-84th  interval of quenched fractions in the PC extended regions (masking out the PC cores), across the 14 different simulations, independently of their host halo mass. Results are shown for galaxies with stellar masses in the range $ \rm 10^9<M_{\ast}/M_{\odot}<10^{10}$ and $ \rm M_{\ast}>10^{10} \ M_{\odot}$ separately.}
    \label{fig:qf_m}
\end{figure}
\newline
We investigate the effects of the local environment, by isolating galaxies hosted by PC cores of different halo masses and galaxies populating the extended PC structures. We also separate galaxies in two stellar mass bins ($\rm 10^9<M_{\ast}/M_{\odot}<10^{10} $ and $ \rm M_{\ast}>10^{10} \ M_{\odot}$) to examine how galaxies of different masses are differently affected by the environment. The median quenched fractions as a function of halo mass, and the fractions in the wide PC structures and in the cosmological box, are shown in Fig. \ref{fig:qf_m}. 
\newline
Across all stellar masses, the quenched galaxy fraction increases significantly in more massive halos, reaching up to $90\%$ ($70\%$) for the $ \rm 10^{9}<M_{\ast}/M_{\odot}<10^{10}$ ($ \rm M_{\ast}>10^{10} \ M_{\odot}$) galaxy population at $ \rm M_{200c}\ge10^{14} \ M_{\odot}$. The scatter around the median is representative of the halo-to-halo variation, though higher halo mass bins  might be affected by limited statistics.
\newline
The {\tt DIANOGA} simulations predict an increased efficiency of quenching mechanisms compared to the cosmological box, across all the probed stellar mass and halo mass ranges, already at $z=2.2$.  Noticeably, the fractions of quenched galaxies in the extended PC regions exceed those in the cosmological box, suggesting that star formation in galaxies might be suppressed before their infall into the virial region of a massive halo. 
\newline
While these simulations predict an increasing quenched fraction as a function of stellar mass in the cosmological box and in the extended PC regions, in qualitative  agreement with field observations \citep[e.g.,][]{muzzin13,kawinwanichakij17,sherman20,park24}, we note that Fig. \ref{fig:qf_m} shows instead higher quenched fractions for $ \rm 10^9<M_{\ast}/M_{\odot}<10^{10}$ than for $ \rm M_{\ast}>10^{10} \ M_{\odot}$ galaxies  in  PC cores, at odds with observations in the field, as mentioned above, as well as in observed clusters and PCs \citep[e.g.,][]{vanderburg13,cooke16,leebrown17,vanderburg20,edward24}.
This issue is common in various cosmological simulation suites across different redshifts, stellar masses, and environments \citep[e.g.][]{bahe17, donnari21}. This discrepancy is likely a direct consequence of the over-consumption of gas in low-mass star-forming galaxies, as discussed in Sect. \ref{sec:ms}, and will be further investigated in a forthcoming study.
\begin{figure}
    \centering
    \includegraphics[width=1\linewidth]{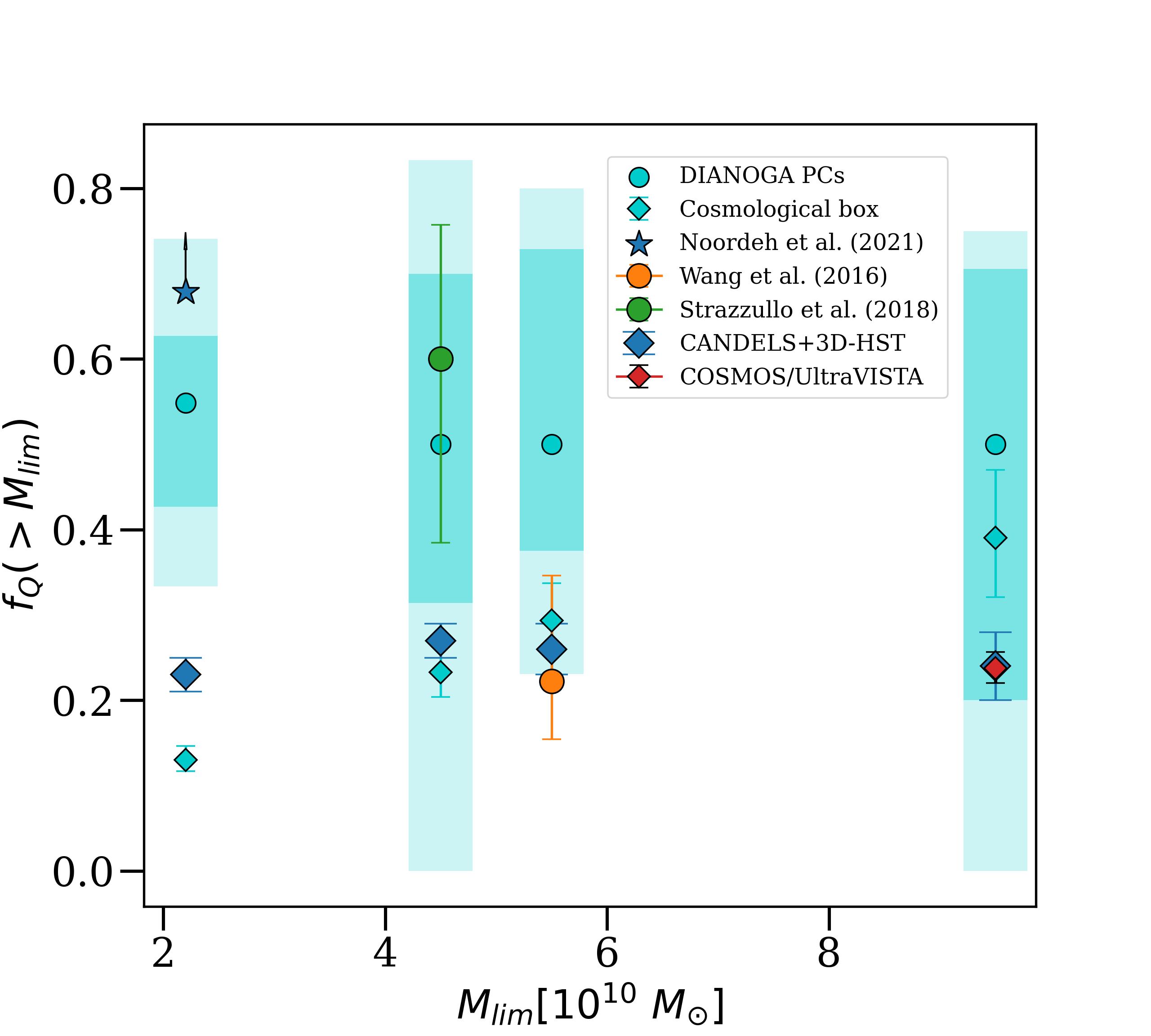}
    \caption{Comparison of quenched fractions in the most massive simulated PC cores (with $\rm M_{200c}\geq 5 \times 10^{13} \ M_{\odot}$) and observations at $z\sim2-2.5$.  Note that measurements in this figure are not homogeneous in terms of the size of the considered region around the PC center (see text).  The cyan dots show the median quenched fractions among selected halos, while shaded dark (light) regions are the 16th-84th percentile (minimum-maximum) ranges. The cyan diamonds, with binomial population uncertainties \citep{cameron11}, are quenched fractions extracted from the cosmological box. For comparison with the observations in CLJ1001, XLSSC 122 and ClJ1449, galaxies in PCs were selected above a limiting stellar mass ($ \rm M_{lim}$) and within a physical aperture around halo centers, matching those adopted in \cite{wang16}, \cite{noordeh21} and \cite{strazzullo18}, respectively (see text for details). For comparison with the cosmological box, we also show the quenched fractions in the field from COSMOS/UltraVISTA catalog \citep{muzzin13} and from CANDELS/3D-HST survey \citep{skelton14} at $2<z<2.5$.}
    \label{fig:qf}
\end{figure}
\newline
In Fig. \ref{fig:qf} we compare the quenched galaxy fraction in the most massive protocluster cores, with $ \rm M_{200c}>5 \cdot 10^{13} M_{\odot}$, with observations in similar environments at $z\sim2-2.5$, including  XLSSC 122 \citep{willis20,noordeh21}, CLJ1001  \citep{wang16} and ClJ1449 \citep{gobat11,strazzullo18}. We note that, while the quoted quenched fractions for XLSSC 122 and CLJ1001 are computed in very similar apertures (approximately 600 and 700 kpc, respectively) and the two (proto)clusters have similar halo masses ($ \rm M_{200c}\sim8-9 \times 10^{13} \ M_{\odot}$), the quoted quenched fraction in ClJ1449 (with $ \rm M_{200c}\sim5\times10^{13} \ M_{\odot}$) is computed within a much smaller aperture of 200 kpc. 
To compare the quenched fractions in simulated PCs with these observations, we selected galaxies in the same apertures and with stellar masses larger than the completeness limits in the referenced observational studies.  We thus stress that measurements in different stellar mass bins shown in Fig. \ref{fig:qf} are not homogeneous in terms of the size of the considered region around the PC centers.  
We also compare the quenched fractions in the cosmological box with field observations in COSMOS/UltraVISTA \citep{muzzin13} and in CANDELS/3D-HST \citep{skelton14} at $z\sim2-2.5$.  For completeness, we also show the quenched fractions in the PC cores in the highest stellar mass bin, even though there is no direct comparison with observed PCs. We calculated these quenched fractions, as a reference, within the average $ \rm R_{200c}$ (440 kpc) of the considered halos. 
The massive halos in the simulated PCs exhibit a broad range of quenched galaxy fractions, reflecting the diversity in the timing of evolutionary processes affecting galaxies in overdense environments at cosmic noon. In particular, the most star-forming among massive PC cores show fractions of quenched galaxies comparable to those in the cosmological box as do observed PCs with negligible environmental quenching \citep[e.g., CLJ1001, ][]{wang16}, while the most quiescent PC cores resemble the most mature observed systems \citep[e.g., XLSSC 122,][]{noordeh21}\footnote{We estimate the mass completeness limit based on the spectroscopic completeness limit from \cite{willis20}, assuming a BC03 SSP with formation redshift $z=5$ and no dust.}. Quenched fractions in the cosmological box in the stellar mass range probed in Fig. \ref{fig:qf} are close to observed quenched fractions in the field at similar redshift, especially at $ \rm M_{\ast}\sim 4-5\times10^{10} \ M_{\odot}$, whereas at higher (lower) masses they are slightly overestimated (underestimated).
\newline
Based on results from these simulations, the large scatter observed in the quenched fraction of PC cores is primarily driven by intrinsic variability in the evolutionary phase of PC cores of similar mass at $z\sim2$. We also note that the quenched fractions of galaxies in the innermost regions of massive PC cores, as shown in comparison with ClJ1449, have the largest scatter. While the intrinsic variability might indeed be enhanced in the selected innermost region of massive halos, the increased scatter might also be due to lower statistics in such small volumes. A more detailed investigation of the processes affecting galaxy evolution in halos of different masses and at different halo-centric radii will be presented in Esposito et al. (in prep.).

\subsection{UVJ diagrams}
\label{sec:uvj}
UVJ diagrams are widely used to classify galaxies as star-forming or quenched in observations \citep[e.g.,][]{labbe05,williams09,whitaker12}. \cite{donnari19} found that the fractions of quenched galaxies inferred from UVJ diagrams in the IllustrisTNG simulations were consistent with those estimated from the galaxy distance from the MS at redshifts $0 \leq z \leq 2$, by adopting a separation line between star-forming and quenched galaxies that preserves the slope of the commonly used classification  \citep[e.g., ][]{whitaker11}, but is shifted to account for the slightly different color distribution compared to observations. We test this approach in the {\tt DIANOGA} simulations, coupled with the SKIRT-9 radiative transfer code (see Sect. \ref{sec:skirt}), to investigate potential complications in the comparison between galaxy populations in simulations and observations, particularly in the context of high-redshift massive halos. 
\newline
\begin{figure*}
 \centering
 \includegraphics[width=20cm]{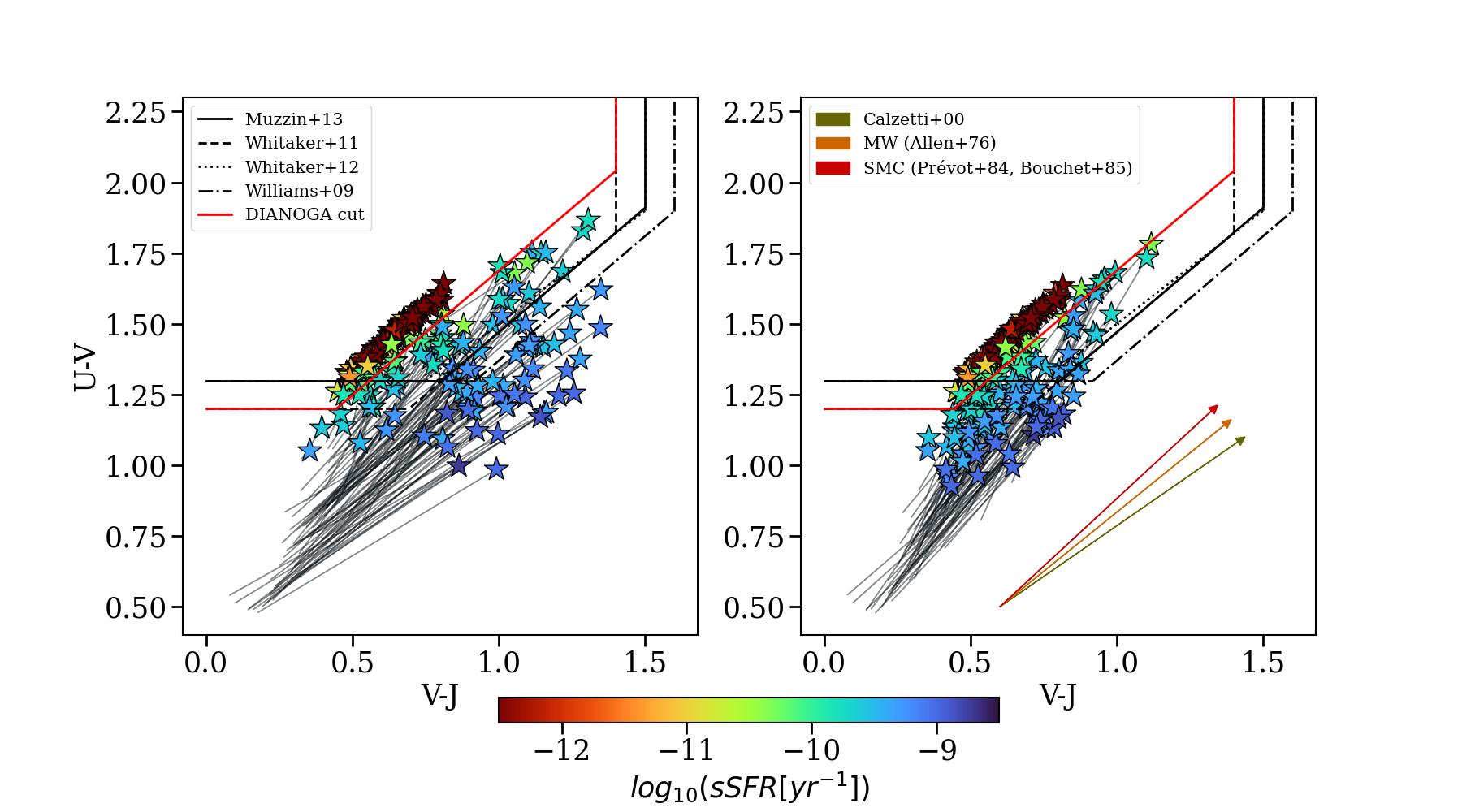}
 \caption{UVJ diagrams for galaxies within 7 massive halos in {\tt DIANOGA} PC regions at $z=2.2$, within a 700 kpc aperture, produced with SKIRT-9. Intrinsic colors are connected to attenuated colors by a grey arrow, ending with a star symbol color-coded according to the sSFR of each galaxy. The left panel shows the diagram obtained by tracing dust in the diffuse ISM through a fixed dust-to-metal ratio. The right panel shows the diagram produced with the model for MCs described in Sect. \ref{sec:skirt}. Typical cuts adopted in UVJ diagrams to separate quenched and star-forming galaxies \citep{williams09,whitaker11,whitaker12,muzzin13} are shown as black lines, while the red line defines a custom-defined boundary to separate these populations in the {\tt DIANOGA} simulations. The colored arrows show the reddening vectors for typical attenuation laws \citep{allen76,prevot84,bouchet85,calzetti}.}
 \label{fig:uvj}
\end{figure*}
We produce UVJ diagrams for a random sample of seven massive halos, with $ \rm M_{200c}=(0.6-1.4)\times10^{14} \ M_{\odot}$, in our simulations. We select galaxies with $ \rm M_{\ast}>2\times10^{10} \ M_{\odot}$, which is approximately the mass range that can be reasonably probed with mass-complete samples in deep observational investigations of z$\sim$2 protocluster galaxy populations thus far. We select galaxies located within a 3D aperture radius of 700 kpc  from the center of each halo, corresponding to $ \rm \sim 1.5  \times R_{200c}$. The resulting UVJ diagrams are shown in Fig. \ref{fig:uvj}. The two panels refer to results based on two models, corresponding to including or not a prescription for MCs (see Sect. \ref{sec:skirt}). In the following, we will refer to these two models as ``diffuse model'' (left panel) and ``MC model'' (right panel), respectively. We remind that the diffuse model relies on a smaller number of assumptions and represents a more direct prediction of the simulations, albeit not resolving the scales of MCs.
\begin{figure*}
 \centering
  \begin{subfigure}{\textwidth}
  \includegraphics[width=\textwidth]{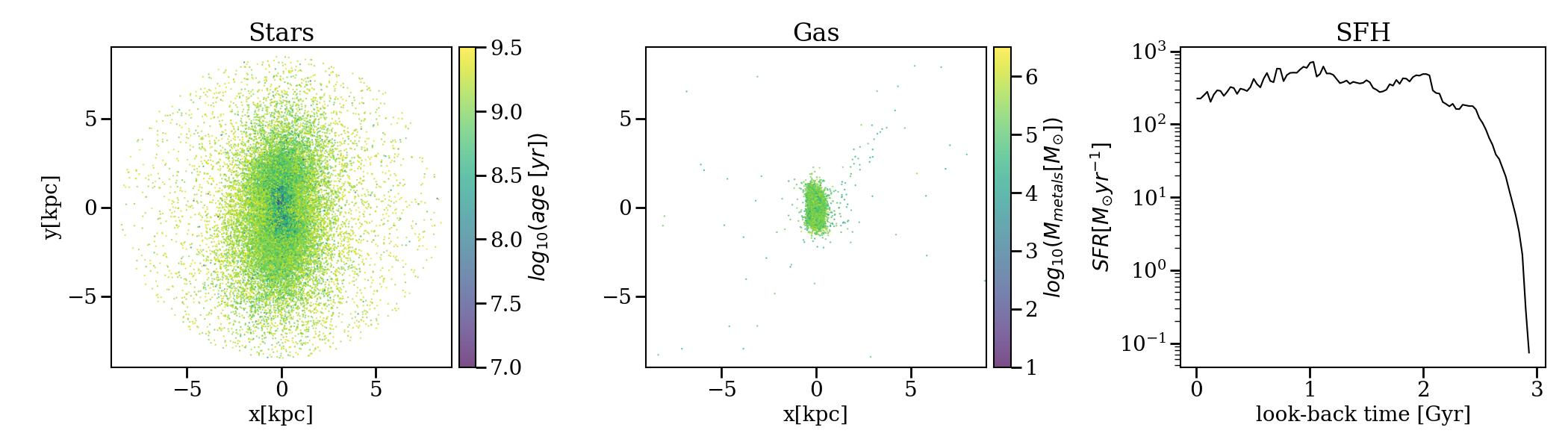}
 \end{subfigure}
  \caption{Properties of a galaxy characterized by a steep reddening vector. The left panel shows the distribution of stars, color-coded according to their age, while the middle panel shows the distribution of gas particles color-coded according to their mass in metals, both within 0.25 kpc from the center of the galaxy along the line of sight. The right panel shows the galaxy star formation history.}
  \label{fig:cfr_outlier}
\end{figure*}
\newline
The diagrams for both models reveal that galaxy populations in the simulated halos exhibit a red sequence \citep[e.g., ][]{bell04} which is already in place at $z=2.2$, reflecting the high quenched fractions presented in Sect. \ref{sec:qf}. These red galaxies form a tight sequence in the quenched region of the diagram and are largely unaffected by dust attenuation, due to the limited gas reservoir in quenched galaxies, which translates into little dust to absorb the stellar emission in our modeling. Qualitatively, the distribution of both quenched and star-forming galaxies is in agreement with observed UVJ diagrams. However, the actual color distribution of star-forming simulated galaxies is not the same as for observed counterparts, therefore, to quantitatively match the separation between the two galaxy populations  based on their sSFR (see Fig. \ref{fig:ssfr_hist}), we need to shift the separation line by approximately 0.2 mag towards redder U-V colors compared to standard separations used in the literature \citep[e.g.,][]{williams09,whitaker11,whitaker12,muzzin13}. We note that this is in quantitative agreement with the UVJ separation found by \cite{donnari19} in the IllustrisTNG simulations, as well as with the color distributions in SIMBA simulations at $z=2$ from \cite{akins22}.
\newline
For star-forming galaxies, the diffuse model produces overall stronger attenuation (the average $ \rm A_V$ for the studied sample is 1, compared with 0.5 for the MC model), that is dominated by the attenuation of the diffuse dust on the more abundant older stellar population. In this respect, we recall that the MC model subtracts the cold gas reservoir from the total gas contributing to the attenuation of older stars in the ISM. Since star-forming galaxies in PCs reach very high fractions of cold to total gas, this leaves overall a rather small amount of diffuse gas, and consequently dust, in the MC model, resulting in lower dust attenuation compared to the diffuse model. 
The details of this effect are related to the assumptions adopted in the MC model implemented here, however, as discussed in Sect. \ref{sec:skirt}, tuning this model is not the focus of this analysis; our primary interest in modeling the age-dependent attenuation of MCs is understanding its qualitative influence on the reddening of galaxy colors in the UVJ diagrams.
\newline
Figure  \ref{fig:uvj} shows that, in our modeling of dust attenuation applied to {\tt DIANOGA} simulations (diffuse model), the majority of the galaxy population has reddening vectors in agreement with typical attenuation laws \citep[e.g.,][]{allen76,prevot84,bouchet85,calzetti}. However, a subset of star-forming galaxies exhibit steeper reddening vectors. This is consistent with the results from \cite{akins22} on SIMBA simulations coupled with dust radiative transfer. In fact, they found that the assumption of a universal \cite{calzetti} attenuation law might impact the efficacy of the UVJ classification, especially for star-forming galaxies near the quenched region of the UVJ diagram. This leads to the incorrect classification of these galaxies as quenched when applying standard UVJ cuts \citep{williams09,whitaker11,whitaker12,muzzin13}. This behavior appears in both our models. Thus, although the MC model seems to extend this feature to a larger fraction of the galaxy population compared to the diffuse model, this is not a numerical artifact introduced by our modeling. We note that galaxies characterized by such ``atypical'' steep reddening vectors tend to be centrals, or anyway located very close to the central galaxies, and have high stellar masses ($ \rm \gtrsim 10^{11} \ M_{\odot} $). The comparison between the two models suggests that the limited resolution of cosmological simulations, which cannot capture the high densities typical of molecular clouds, does not explain this inconsistency in the simulated UVJ diagram. Indeed, if anything, the MC model produces a more pronounced tilt of the reddening vector in the UVJ plane.
\newline
We further investigate the ``atypical'' galaxies by analyzing the spatial distributions of their stars and metals (whose mass is mapped into dust mass in our modeling) and their star formation histories. As an example, we show in Fig. \ref{fig:cfr_outlier} these quantities for one of these galaxies. The star formation history (right panel) shows approximately constant SFR for the last $\rm \sim 2 \ Gyr$ at $ \rm \sim 400 \ M_{\odot} yr^{-1}$. The distribution of stars (left panel) shows a positive gradient of stellar ages, while the distribution of metals (central panel) is concentrated in a dense central core. The steepness of the reddening vector is thus produced by the presence of a central population of young stars, that are highly attenuated by the core of high-metallicity gas. This is a genuine signal of age-dependent extinction produced by the simulations (though on larger scales than the unresolved MCs), that causes a stronger reddening of the U-V color than of the V-J color (see Fig. \ref{fig:uvj}).
\section{Summary and conclusions}
\label{sec:conclusions}
In this study, we have analyzed the properties of galaxy populations in protoclusters at $z=2.2$ in the {\tt DIANOGA} set of zoom-in cosmological hydrodynamical simulations of galaxy clusters. These include 14 re-simulations of galaxy clusters originally selected in a DM-only simulation box of size 1 cGpc/h. Alongside the protocluster simulations, we also analyzed a cosmological box of 49 cMpc/h per side as representative of an average region of the Universe, simulated at the same resolution and including the same galaxy formation model in terms of sub-resolution descriptions of the relevant astrophysical processes.
\newline
To define protoclusters in simulations, we traced back the DM that falls within the $ \rm R_{200c}$ of the $z=0$ cluster and identify, around the most massive progenitor halo, a region that encompasses $80\%$ of all these DM particles. In all our analyses we considered galaxies with $ \rm M_{\ast}>10^9 \ M_{\odot}$, so as to ensure an adequate numerical resolution.
\newline
We investigated the properties of galaxy populations in the simulations, and compared to observations to test their predictions, focusing on halos with similar properties to the observed (proto)cluster cores. We compared galaxy properties in the simulated protoclusters and in the cosmological box to highlight the impact of environment on galaxy evolution, as predicted by our simulations. We also performed radiative transfer simulations using the SKIRT-9 code \citep{skirt} to estimate dust-attenuated rest-frame UV to NIR colors of simulated galaxies consistently with their gas and metallicity properties, and test the impact of adopting a UVJ photometric classification of star-forming and quenched galaxies, as routinely applied in observational studies. 
\newline
We summarize our main conclusions as follows.
\begin{itemize}
\item The galaxy stellar mass function (GSMF) in the {\tt DIANOGA} protoclusters is broadly consistent with observations in protoclusters at $2< z \leq 2.5$, both in shape and normalization (see Fig. \ref{fig:gsmf-obs} ). This demonstrates that the integrated star formation history in the simulated protoclusters realistically reproduces the observational data. The median protocluster GSMF reveals an excess of galaxies with $ \rm M_{\ast}>10^{10} \ M_{\odot}$ compared to the cosmological box (see Fig. \ref{fig:gsmf}). This excess is detected both in protocluster cores ($ \rm M_{200c} \gtrsim 1.5 \times 10^{13} M_{\odot}$) and, albeit with lower significance, in the extended protocluster regions (i.e., outside the cores). Since this excess is indicative of early stellar mass assembly, its presence in the extended protocluster regions suggests the onset of an overall accelerated galaxy evolution in the broader environments hosting protoclusters.
\item The star formation rates (SFRs) of star-forming galaxies in both the protoclusters and the cosmological box are lower than expected from observations (see Fig. \ref{fig:ms}), consistent with previous results from lower-resolution versions of the {\tt DIANOGA} simulations \citep{granato15,bassini20} and other suites of cosmological hydrodynamical simulations \citep[e.g.,][]{akins22,edward24}.
SFRs in protocluster galaxies show a small ($\sim 0.05$ dex) yet significant suppression compared to galaxies in the cosmological box (see Fig. \ref{fig:ms_offset}). This signal is dominated by galaxies in the extended structures of the protoclusters, while galaxies in the protocluster cores exhibit a more pronounced suppression (up to $\sim 0.25$ dex lower SFRs compared to the cosmological box). 
\item The reservoirs of cold star-forming gas are systematically lower than observational estimates at $z \sim 1.5 - 3$, while depletion times are broadly consistent with those observed in star-forming galaxies, both in the protoclusters and in the cosmological box (see Fig. \ref{fig:sfr_drivers}). This suggests that the underestimation of SFRs in the simulations is related to insufficient gas available to be converted into stars, possibly due to the effect of a too efficient feedback inhibiting cooling or expelling gas.
\newline
Protocluster galaxies in our simulations have marginally lower cold gas content than their counterparts in the cosmological box on average, while depletion times do not show significant differences (see Fig. \ref{fig:sfr_drivers_offsets}). However, galaxies in protocluster cores exhibit more marked differences in both cold gas fractions and depletion times compared to galaxies in the extended protocluster structure and in the cosmological box. In massive cores, both cold gas fractions and depletion times show a significant drop for galaxies with $ \rm log_{10}(M_{\ast}) < 9.5$, and remain significantly lower than the cosmological box levels at higher masses. This is due to a large fraction ($\sim 40\%$) of galaxies with $ \rm log_{10}(M_{\ast}) < 9.5$, that are classified as star-forming based on their SFR averaged over the past 200 Myr, but have actually completely lost their cold gas over this time.
\newline
We note that these galaxies may be in a regime influenced by the finite numerical resolution. However, since galaxies of similar mass in other simulated regions do not show such a sudden loss of gas, this feature could reflect a genuine physical process emerging in simulated massive halos at high redshift, possibly amplified by the combined effect of relatively limited numerical resolution and the extreme environment of the most massive high-$z$ halos.
\newline
\item The fraction of quenched galaxies in the simulations increases significantly with the mass of the host halo in the protocluster cores, reaching approximately 0.9 (0.7) for galaxies with $ \rm 10^9 < M_{\ast}/M_{\odot} < 10^{10}$ ($ \rm M_{\ast} > 10^{10} \ M_{\odot}$, respectively; see Fig. \ref{fig:qf_m}). 
Our simulations predict an increasing fraction of quenched galaxies towards lower stellar masses in protocluster cores, at odds with indications from observations \citep[e.g.,][]{muzzin13,vanderburg13,cooke16,leebrown17,kawinwanichakij17,vanderburg20,sherman20,park24,edward24}, and also with our predictions in the cosmological box and outside the cores of the simulated protoclusters. This trend is common to many suites of cosmological simulations \citep[e.g.,][]{bahe17,donnari21}. The fraction of quenched galaxies in protoclusters, even outside massive cores, is higher than in the cosmological box, implying accelerated quenching also in lower-density PC regions, before infall into massive halos. 
\item {\tt DIANOGA} protoclusters include halos in which quenching has progressed in very different ways (see Figs. \ref{fig:qf_m},\ref{fig:qf}), similarly to observed PCs. In some halos, galaxy populations are as star-forming as in the cosmological box, comparable to observed starbursting protocluster cores \citep[e.g., CLJ1001 at $z = 2.51$,][]{wang16}, while in others are as quenched as observed mature clusters \citep[e.g., XLSSC 122 at $z = 1.98$,][]{willis20,noordeh21}. The quenched fractions in the cosmological box also qualitatively agree with observations in the field. 
\item We found that the distribution of galaxies in the simulated UVJ diagram in a sample of 7 massive simulated halos is qualitatively consistent with observed distributions of star-forming and quenched galaxies (see Fig. \ref{fig:uvj}).  However, from a more quantitative point of view, we needed to shift the UVJ selection boundary by 0.2 mag toward redder U-V colors, compared to typical classification criteria used in observations, to optimally match the sSFR-based classification of our simulated galaxies. Despite the overall qualitative agreement, we note indeed the presence of a population of star-forming galaxies (mainly massive galaxies in the central regions of protocluster cores) with a steep reddening vector, leading to significant contamination of the quenched region of the diagram when adopting standard selections. \end{itemize}
In this work, we have demonstrated that the {\tt DIANOGA} cosmological hydrodynamical simulations can broadly reproduce the properties of galaxy populations in observed protoclusters at $z\sim2-2.5$, a particularly significant time for the evolution of clusters and their host galaxies, corresponding to the transition between protocluster structures and the first established clusters, at a cosmic epoch where the cosmic star formation rate density is at its peak and, at the same time, the first signatures of environmental quenching appear in the densest environments.  We show that the environment of cluster progenitors has an impact on galaxy populations, particularly through the early onset and subsequent suppression of star formation, most significant in the massive protocluster cores. These results provide an interpretative framework for understanding the evolution of galaxies in protoclusters, which is driven by both environmental and internal processes, consistently accounting for the full cosmological context.
\newline
The next crucial step is to constrain when, where and how the environment of (proto)clusters affects the evolution of galaxy populations, considering the broad range of environments that these galaxies cross throughout their evolutionary history. 
This includes investigating the build-up of the diffuse Intra-Cluster Medium that permeates mature clusters, as well as examining internal processes within galaxies, including AGN activity. 
\newline
The results of our analysis highlight that the current generation of cosmological hydrodynamical simulations are invaluable instruments to shed light on the complex interplay between environment and the physical processes that shape galaxy evolution. In this respect, cluster progenitor environments at cosmic noon present a significant challenge to our modeling of star formation and feedback processes, coupling extreme regions of the Universe and a cosmic time where transformational processes are at their peak. In this context, and while keeping in mind the current limitations still affecting the comparison of observations and simulations at high redshift, our findings highlight the need for improved modeling of galaxy formation and evolution to better capture observed galaxy population properties - in particular the level of star formation - in the densest high-redshift regions. This is a crucial part of the broader effort to consistently reproduce the evolution of galaxy populations across cosmic time and across different environments.


\begin{acknowledgements}
 We thank Luca Tornatore for technical support, Alice Damiano for useful discussion, and an anonymous referee for constructive comments that helped improve the presentation of the results. Simulations have been carried out at the CINECA Italian Supercomputing Center, with computational resources allocated through an ISCRA-B proposal. This paper is supported by: the Italian Research Center on High Performance Computing, Big Data and Quantum Computing (ICSC), project funded by European Union - NextGenerationEU - and National Recovery and Resilience Plan (NRRP) - Mission 4 Component 2, within the activities of Spoke 3, Astrophysics and Cosmos Observations; by the PRIN 2022 PNRR project (202259YAF) ``Space-based cosmology with Euclid: the role of High-Performance Computing.''; by the PRIN 2022 (20225E4SY5) ``From ProtoClusters to Clusters in one Gyr''; by the INAF Astrofisica Fondamentale GO Grant 2022 ``Environmental quenching efficiency across the virial volume of the most massive, most distant clusters''; by the INAF Astrofisica Fondamentale Large Grant 2023 ``Witnessing the Birth of the Most Massive Structures of the Universe''.  We acknowledge partial financial support from the INFN Indark Grant, from the Consejo Nacional de Investigaciones Cient\'ificas y T\'ecnicas (CONICET) (PIP-2021-11220200102832CO), from the Agencia Nacional de Promoción de la Investigación, el Desarrollo Tecnológico y la Innovación de la Rep\'ublica Argentina (PICT-2020-03690), and from the European Union's HORIZON-MSCA-2021-SE-01 Research and Innovation Programme under the Marie Sklodowska-Curie grant agreement number 101086388 - Project (LACEGAL)
\end{acknowledgements}

\bibliographystyle{aa}
\bibliography{references}

\begin{appendix}
\section{Impact of apertures on stellar masses of simulated galaxies}
\label{sec:app1}
\begin{figure}
    \centering
    \includegraphics[width=1\linewidth]{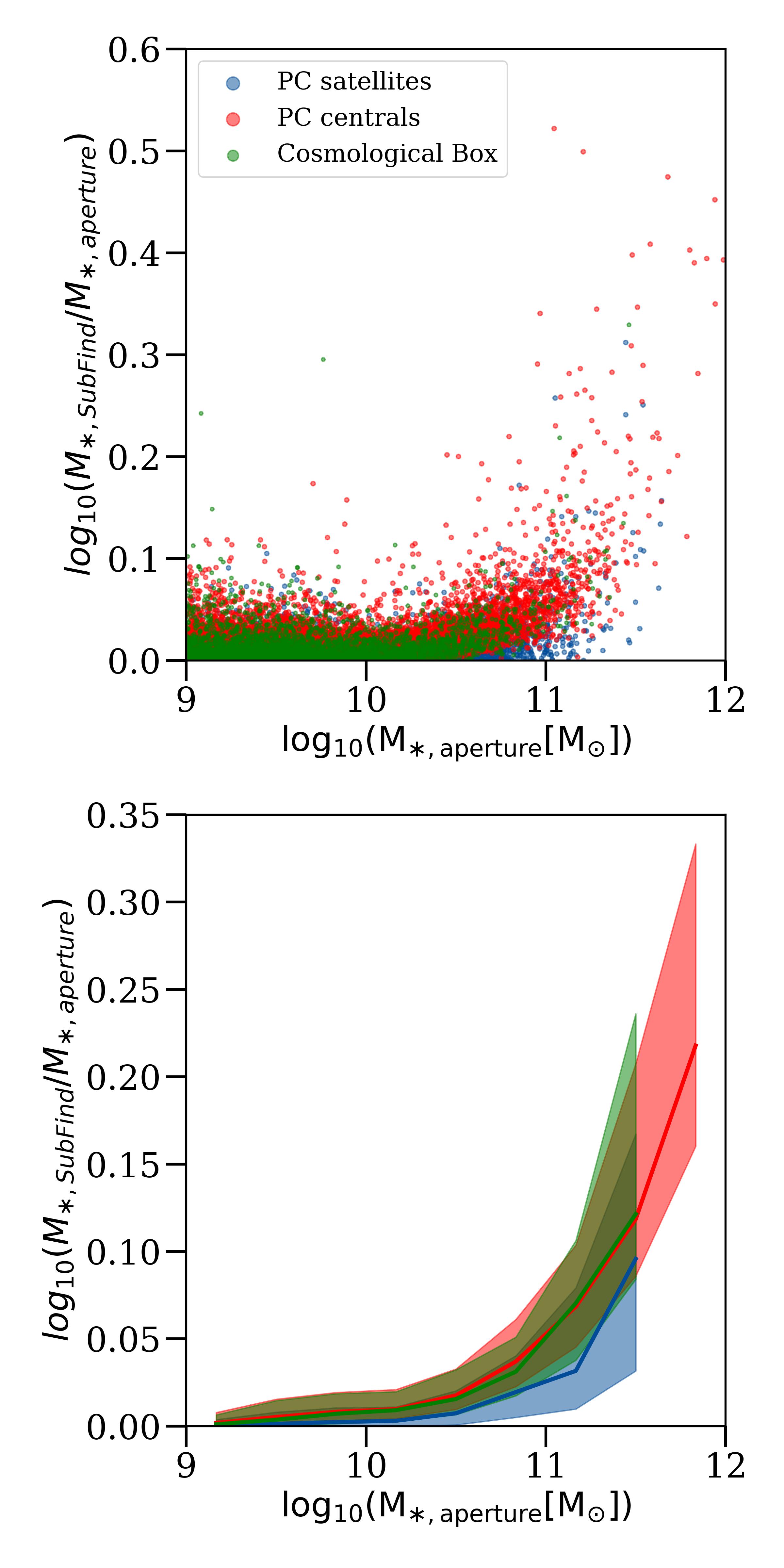}
    \caption{Ratio of {\tt SubFind} and aperture stellar masses, as a function of aperture mass for galaxies in {\tt DIANOGA} PCs, separated in centrals and satellites, and in the cosmological box at $z=2.2$. Aperture masses are estimated within a two-dimensional projected radius equal to 8.5 kpc. The bottom panel shows the median (16th-84th percentiles) of the mass ratios of galaxies in the top panel, in solid lines (shaded areas, respectively).}
    \label{fig:cfr_sf_aperture}
\end{figure}

Here we examine the impact of using stellar masses calculated within fixed apertures (2D, with an 8.5 kpc radius, corresponding to 1 arcsecond at $z = 2.2$) as opposed to using total {\tt SubFind} masses. Fig. \ref{fig:cfr_sf_aperture} shows the ratio of these two mass definitions as a function of aperture stellar mass. This ratio is $\sim1$ on average for stellar masses up to $ \rm M_{\ast} \sim 10^{11} \ M_{\odot}$, with negligible scatter ($\sim 0.005$), for galaxies both in the PCs and in the cosmological box. For $ \rm M_{\ast} > 10^{11} \ M_{\odot}$, however, the deviation of the mass ratio from unity becomes significant, with the {\tt SubFind} masses exceeding the aperture masses by a factor $\sim1.3$ ($1.2$) on average in the PCs (cosmological box, respectively).
\newline
Most galaxies (72$\%$) with $ \rm M_{\ast} > 10^{11} \ M_{\odot}$ are central galaxies in massive halos, where {\tt SubFind} fails to separate the galaxy from the diffuse stellar component \citep[see, e.g.,][]{marini21}. For $ \rm M_{\ast}>10^{11} \ M_{\odot}$ satellites (which are thus not affected by this effect) the {\tt SubFind} mass estimate still exceeds the aperture mass by a factor 1.1 on average, reaching up to a factor $\gtrsim$2 for $1 \%$ of these most massive galaxies. 

\section{Effects of averaging SFRs}
\label{sec:app2}
\begin{figure*}
    \centering
    \includegraphics[width=1\linewidth]{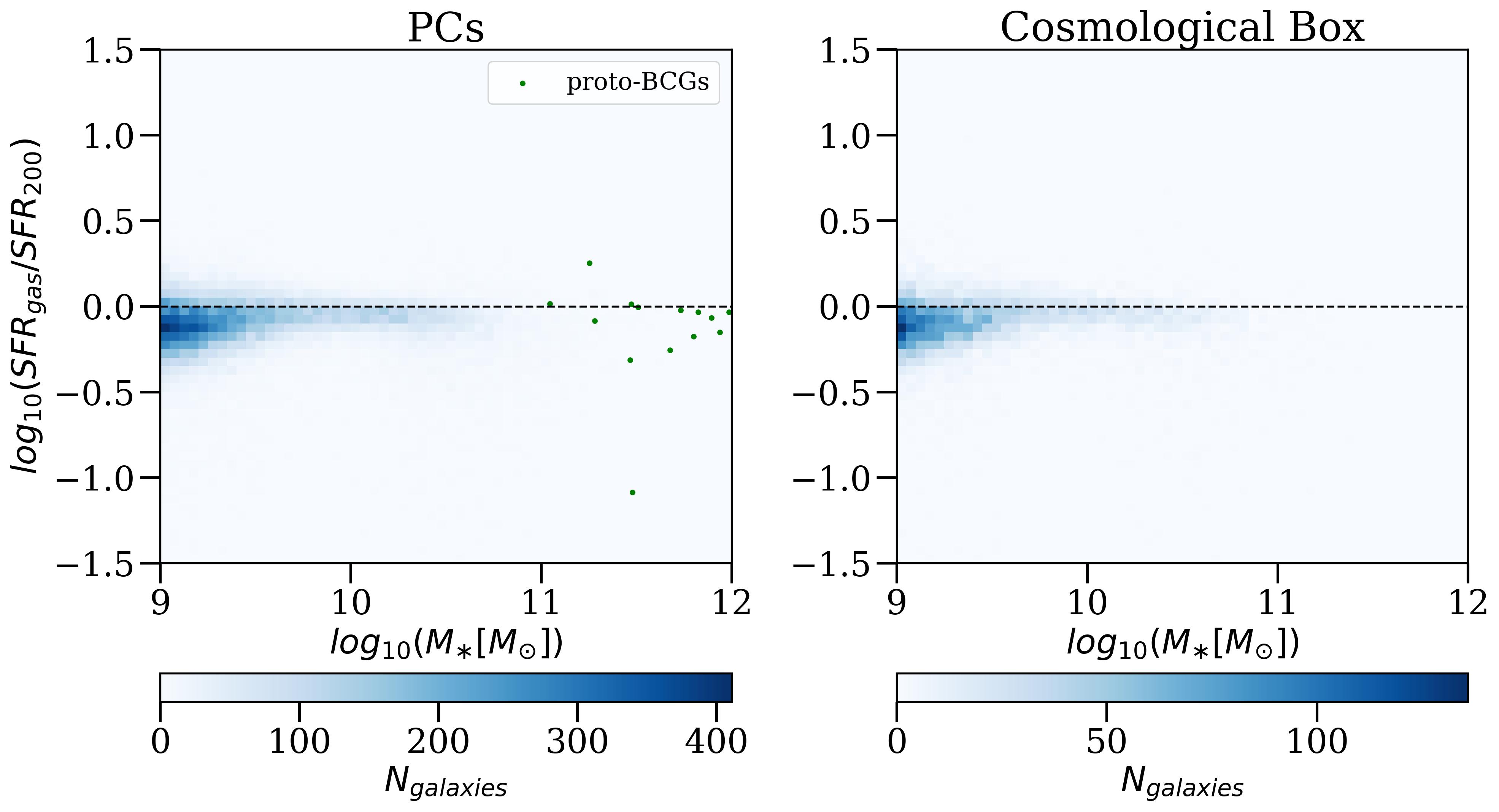}
    \caption{Two-dimensional histograms of the ratios of instantaneous SFR ($ \rm SFR_{gas}$) and SFR averaged over 200 Myr ($ \rm SFR_{200}$) of galaxies in {\tt DIANOGA} PCs (left) and in the cosmological box (right) at $z=2.2$, as a function of stellar mass. Proto-BCGs are plotted separately.}
    \label{fig:cfr_sfrs}
\end{figure*}

\begin{figure*}
    \centering
    \includegraphics[width=1\linewidth]{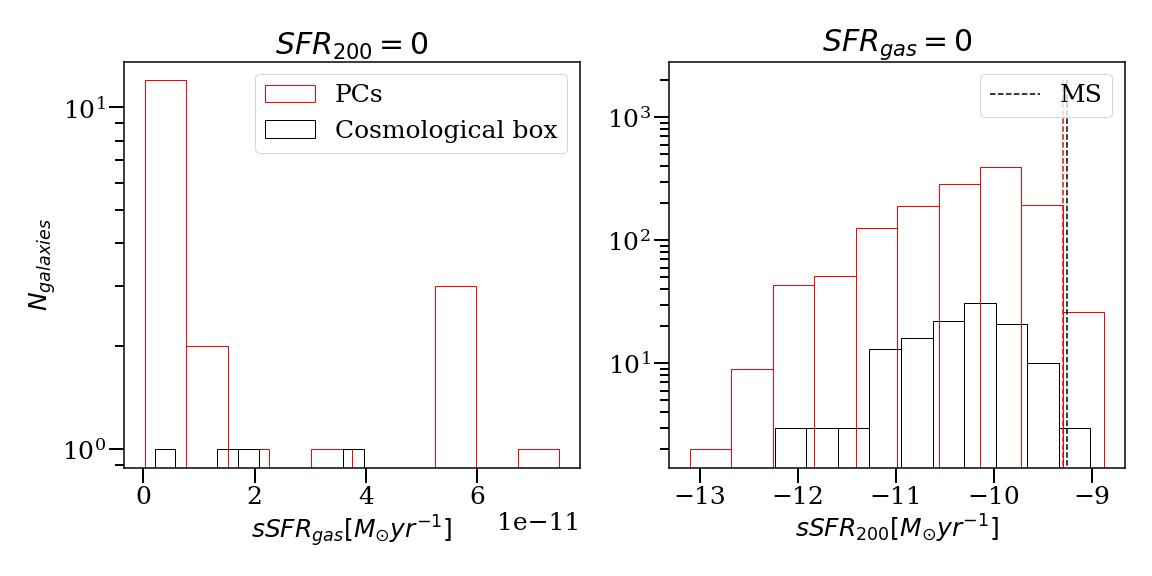}
    \caption{Distribution of the instantaneous (averaged) sSFRs for galaxies with $ \rm SFR_{200} = 0$ ($ \rm SFR_{gas} = 0$), in the left (right, respectively) panel. The dashed lines show the average sSFR of main-sequence galaxies in the cosmological box (black) and in the PCs (red).}
    \label{fig:hist_sfr0}
\end{figure*}

Here we assess differences between galaxy SFRs averaged over 200 Myr ($\rm SFR_{200}$), which more properly compare to some observational estimates used in this work (see Sect. \ref{sec:ms}), and the instantaneous SFRs associated with gas particles in the SH03 star formation model ($ \rm SFR_{gas}$). Fig. \ref{fig:cfr_sfrs} shows the ratio between these two SFR estimates as a function of stellar mass. $ \rm SFR_{gas}$ is lower than $\rm SFR_{200}$ across all the probed stellar mass range, with a weak increasing trend with increasing mass. Proto-BCGs behave similarly to other galaxies, with instantaneous SFRs tendentially lower than those averaged over the past 200 Myr. The same trends are observed in PCs and in the cosmological box, reflecting a declining star formation history of galaxies in both environments at $z=2.2$, possibly more pronounced at lower masses.
\newline
Figure \ref{fig:hist_sfr0} shows the distributions of  specific  $ \rm SFR_{gas}$ ($ \rm SFR_{200}$) for galaxies in which the $ \rm SFR_{200}$ ($ \rm SFR_{gas}$, respectively) is zero, which are not accounted for in Fig. \ref{fig:cfr_sfrs}. There are very few galaxies with $ \rm SFR_{200} = 0$ and $ \rm SFR_{gas} >0$, and their  $ \rm SFR_{gas}$ corresponds anyway to a negligible level of residual star formation. 
Galaxies with  $ \rm SFR_{gas} = 0$ but $\rm SFR_{200}>0$ are more common; these galaxies have relatively low (averaged) sSFRs, at the level of the adopted threshold for quenched galaxies and roughly 0.7 dex below the MS.
Galaxies for which both SFR estimates are zero also do not appear in Fig. \ref{fig:cfr_sfrs}. However, they would in principle add to the population of galaxies for which the estimates are the same and, since they are a minority across all stellar masses, they would not significantly change the trend shown in Fig. \ref{fig:cfr_sfrs}.


\end{appendix}

\end{document}